\documentstyle[12pt,epsf]{article} 

\newcommand{\comm}[1]{}   
\newcommand{\gsim}{\raisebox{-0.8ex}{\mbox{$\stackrel{\textstyle>}{\sim}$}}}

\newcommand{\teff}{T_{\rm eff}}

\newcommand{\ncs}{N_{_{\rm CS}}}
\newcommand{\ndcs}{\dot{N}_{_{\rm CS}}}
\newcommand{\ncsmall}{n_{_{\rm CS}}}

\newcommand{\ncsa}{\langle n_{_{\rm CS}}\rangle}

\newcommand{\ncsmt}{\langle n_{_{\rm CS}}(t)\rangle}

\newcommand{\nw}{N_{\rm wind}}
\newcommand{\ndw}{\dot{N}_{\rm wind}}

\newcommand{\esph}{E_{\rm sph}}
\newcommand{\gsph}{\Gamma_{\rm sph}}
\newcommand{\be}{\begin{equation}}
\newcommand{\ee}{\end{equation}}
\newcommand{\eq}[1]{(\ref{#1})}

\begin{document} 
\title{Resonant Amplification of Electroweak Baryogenesis at Preheating} 

\author{J. M. Cornwall, D. Grigoriev\thanks{On leave of absence from
Institute for Nuclear Research of Russian Academy of Sciences}
\thanks{Present address: Dept.~of Mathematical Physics, National
University of Ireland, Maynooth, Co. Kildare, Ireland} ~and
A. Kusenko\thanks{Also at RIKEN BNL Research Center, Brookhaven National Laboratory, Upton, NY 11973}\\{\small Department of Physics and Astronomy, University of
California}\\[-0.8ex]{\small Los Angeles, CA 90095-1547}}
\date{UCLA/01/TEP/10} \maketitle

\begin{abstract} 
 We explore viable scenarios for parametric resonant amplification of
electroweak (EW) gauge fields and Chern-Simons number during
preheating, leading to baryogenesis at the electroweak (EW) scale.  In
this class of scenarios time-dependent classical EW gauge fields,
essentially spatially-homogeneous on the horizon scales, carry
Chern-Simons number which can be amplified by parametric resonance
 up to magnitudes at which unsuppressed topological transitions
in the Higgs sector become possible.  Baryon number non-conservation
associated with the gauge sector and the highly non-equilibrium
nature of preheating allow for efficient baryogenesis.  The requisite
large CP violation can arise either from the time-dependence of a
slowly varying Higgs field (spontaneous baryogenesis), or from a
resonant amplification of CP violation induced in the gauge sector
through loops.  We identify several CP violating operators in the
Standard Model and its minimal extensions that can facilitate 
efficient baryogenesis at preheating, and show how to overcome
would-be exponential suppression of baryogenesis associated with
tunneling barriers.

\end{abstract}

\section{Introduction} 
 
Parametric resonance coupling of the oscillating inflaton to Standard Model
fields \cite{kls} offers new dynamical mechanisms for various
early-universe phenomena.  One interesting scenario \cite{kt,ggks,gg} adjusts
inflaton parameters\footnote{The consistency of low-scale inflation with
primordial density perturbations has been shown by Germ\'an {\it et al}
\cite{grs}.} so that reheating 
does not heat the universe to a temperature above the EW cross-over 
temperature $T_c$.  This means that sphaleron transitions are frozen out
and that B+L created before reheating will not be washed out.
 
The early attempts \cite{krs,ckn} to create B+L solely through Standard
Model effects suffered from the fact that if the only source of CP
violation is in the CKM matrix, it would lead to far too small a value of
B+L, and from the need for a first-order EW phase transition to drive
out-of-equilibrium effects. This first-order transition seems to be ruled
out by current lower limits on the Higgs mass of around 110 GeV.  So it
might appear that EW baryogenesis is unattractive.  However, with the
suggestion of EW preheating and the accompanying parametric resonances, it
has become interesting to look at EW baryogenesis in a different way.
 
This was done in a recent paper \cite{alexmike}, where two of us showed
that EW parametric resonance, with the Higgs field oscillating because of
its coupling to the oscillating inflaton, could amplify CP-violating seed
values of the EW gauge potential in an $ansatz$ where this gauge potential
was spatially-homogeneous (out to the horizon) but time-dependent.  The CP
violation was manifested as a homogeneous classical ``condensate" of
Chern-Simons number or equivalently of B+L (through the EW anomaly).  The
scenario of \cite{alexmike} had several shortcomings:
\begin{enumerate}
\item CP violation only occurred because of CP-violating initial values for
the gauge potential, not because of any explicit CP violation in the gauge
equations of motion.
\item The Higgs field oscillations were considered as given, so that no
(classical) back-reaction of the gauge fields on the Higgs field was
considered.
\item No plausible scenario was suggested for the permanent conversion of
Chern-Simons number to actual baryons and leptons; once parametric
resonance driving ended, it would be possible for the homogeneous
Chern-Simons condensate to dissipate.  Actual formation of baryons and
leptons requires the development of spatial inhomogeneities, rather similar
to sphalerons; one might term this process {\em sphalerization}.
\end{enumerate}   
 
In the present work we extend the considerations of \cite{alexmike} to
deal, at least partially, with these three shortcomings.

 First, we
introduce explicit CP violation\footnote{The CP-violating terms used here
are similar to those invoked in spontaneous baryogenesis but represent
couplings to the gauge sector; see \cite{ckn1} and references therein.  We
do not require a first-order phase transition, as commonly invoked in
spontaneous baryogenesis.  For a version of spontaneous baryogenesis in the
present context, see Sec. \ref{spon} below.} in the EW gauge plus Higgs
equations of motion.  This 
CP violation comes from coupling of CP-violating effects to the EW gauge
fields through loops, typically quark loops.  We explore two cases; in one,
there is out-of-equilibrium strong CP violation as evidenced by an $\eta'$
field slowly (on EW time scales) rolling in a standard QCD potential with
zero $\theta$ angle (so that there is no strong CP violation when the
$\eta'$ field is in equilibrium).  In the other, we invoke
spontaneously-broken CP in multi-Higgs models \cite{lw}, again, out of
equilibrium.\footnote{These multi-Higgs models are not realistic for CP
violation in the $K-\bar{K}$ system, but they could still play a central
role in baryogenesis; note that in general our scenarios for baryogenesis
involve CP violation which is far from the equilibrium values seen today.} 
Note that it is not easy to rule out the CP-violating operators we use by
present-day experimental results on CP violation, since in our scenario the
CP-violating effects are far from their equilibrium values.

Second, we study the combined gauge field-Higgs classical equations,
extending the previous $ansatz$ to include a spatially-homogeneous
time-dependent Higgs field.  The classical backreaction of the gauge potential on the Higgs field can drive the Higgs field into oscillating through zero VEV rather than staying near a broken-symmetry minimum.  This is important for sphalerization.

Third, in the final section of the paper we make some remarks about the dynamics of sphalerization.
This is a truly difficult dynamical problem, requiring extensive
numerical investigation, and we barely scratch the surface here. Aside from
brute-force computation, not attempted here, there are several avenues to
explore for guidance, including simple effective-temperature arguments
previously used in connection with EW preheating\cite{kls,ggks,gg};
numerical studies of the 1+1-dimensional Abelian Higgs
model\cite{ggks,gg,ew2000}; and approximate but useful tools for an
analysis  based on tools developed for understanding the possibility of B+L
violation in high-energy two-particle collisions (see, {\it e.g.},
\cite{mamo}). A major obstacle to the analysis is that mere production of
Chern-Simons number is not enough to produce baryons; the Chern-Simons
number must be converted into baryons through the medium of formation of
Higgs winding number $\nw$. 
Using simple topological arguments, we construct a model for the
energy profile of the system  which illustrates the issues involved.  With
the simplification of using an effective temperature we estimate 
baryoproduction for an arbitrary time dependence of the topological
transition rate  and show how wash-out---fermionic backreaction leading to
dissipation of the newly-created baryons and leptons---can be taken into
account.   Finally we attempt an alternative to effective-temperature
considerations by extending some approximate techniques used long ago to
examine the possibility of B+L violation in two-particle collisions. 

The upshot of these considerations is that there are
at least two possibilities for conversion of Chern-Simons number to
baryons which do not suffer from the exponential suppression of
tunneling.  In the first, various effects (such as gauge backreaction
on the Higgs field) may cause the Higgs VEV $v(t)$ to oscillate
through zero during or immediately after preheating, allowing
unsuppressed transitions (the sphaleron mass $M_S\sim v/g$ vanishes).
In the second, formation of spatial structures on various scales may
allow for baryon formation to proceed at energies above barrier
heights even if $v$ is near its usual broken-symmetry value. In this
case, there is a relation between the size of the structure, the
energy of the structure, and the Chern-Simons density, which we
explore with a relatively crude but qualitatively satisfactory model
of the topological charge barrier factor as a function of energy and
size. We argue that as various spatial scales arise and grow during
preheating and the Chern-Simons condensate becomes inhomogeneous, it
is possible to have conversion of Chern-Simons number to actual
baryons (and leptons, which we ignore in this paper) which is not only
unsuppressed with regard to topological charge tunneling, but is also
not exponentially suppressed by poor overlap between initial and final
states. This is an important point.  About a decade ago, it was
suggested that EW collisions of two particles yielding an $N$-particle
final state at energies $E \geq M_S$ large enough to overcome the
barrier height and with $N\sim 4\pi/g^2$, could lead to unsuppressed B+L
production (see, {\it e.g.}, \cite{mamo} for a contemporaneous
collection of papers on the subject).  But others (including Banks
{\it et al} \cite{mamo} and Cornwall \cite{mamo,co90}) pointed out
that the poor overlap between the initial and final states led to
exponential suppression anyhow, with a barrier which was a finite
fraction of the canonical 't Hooft barrier.  The suppression mechanism
had actually been proposed earlier by Drukier and Nussinov \cite{dn}
to show that production of solitons such as the sphaleron in
two-particle collisions was exponentially suppressed.

In the present case, the initial state for baryogenesis is not a
two-particle plane-wave state, but a Chern-Simons condensate which is
largely homogeneous and on which various spatial ripples are growing,
as suggested in \cite{alexmike}; when the gauge potential grows to be
of order $m/g$, the spatial growth rate is of order $m$.  
We show that this sort of condensate can have good overlap with
baryonic final states (as connected via the EW anomaly) at certain
spatial scales under the same circumstances (generally high enough
energy) where the topological barrier is gone.  
These spatial scales, to no one's surprise, are around $m^{-1}$ at
energies near the sphaleron mass.  
Of course, if the Higgs VEV $v$ is rather small, the corresponding
spatial scale grows larger and the energy scale grows smaller.
We show that if this second mechanism to avoid suppression is to work
when $m$ is near its vacuum value, the original spatially-homogeneous
Chern-Simons condensate must have a density of order $10^{-2}-10^{-3}$
units of topological charge in a volume $m^{-3}$.  
Such values are indeed reached in numerical simulations (this paper
and Ref. \cite{alexmike}).

We cannot, with these crude approaches, begin to quantify the number of baryons produced.  All we can say is that the number of baryons is essentially linear in the strength of the CP-violating operators we discuss, and that we are not aware of any experimental limits which would lead to strengths too small to produce the observed number of baryons, provided that the baryogenesis process is not exponentially suppressed and that further baryon washout at reheating is not strong.  
Given the results of this paper, the unexplored rate-limiting step for baryogenesis is the growth of perturbations at various spatial scales.

For some simple versions of the models we study it is possible to do
some approximate analysis of growth of Chern-Simons number in the
spatially-homogeneous phase.  Generally we find that adding explicit
CP violation to the gauge equations of motion leads, as expected, to
secular (that is, not oscillating with the inflaton or Higgs)
CP-violating terms in the gauge potential and Chern-Simons density.
If the explicit CP violation is correlated on super-Hubble scales, as
might be expected following inflation and which we will assume for
this paper, the secular average will also be correlated on
super-Hubble scales even though initial values of the gauge potentials
might be random from one Hubble domain to the next.  Without explicit
CP violation in the equations of motion, CP violation can only come
from CP-violating initial values of the EW gauge fields; if (as
mentioned in \cite{alexmike}) these are random from one Hubble volume
to the next and on the average CP-symmetric, the baryon number
averaged over the whole universe will be zero and fluctuations will be
too small by the usual factor of $1/\sqrt N$, where $N$ is the number
of Hubble volumes.  (Of course,  CP-violating effects before preheating can be correlated on super-Hubble scales, so that
there can be a
secular average with no explicit CP violation in the equations of
motion.)  We offer one model, somewhat resembling a Brownian ratchet
\cite{ratchet}, in which the secular Chern-Simons average depends not
only on the sign of the explicit CP violation (assumed to have
super-Hubble correlations) but also on the sign of initial conditions,
which we could assume as random and uncorrelated.  Nonetheless, this
model leads to a sort of secular Chern-Simons average, since resonant
growth can occur only for one sign of initial-condition parameters;
otherwise, there is damping.  In this model, however, the secular
average itself wanders slowly on a long time scale and consequently is
somewhat inefficient.  We also offer other models in which the secular
Chern-Simons sign depends only on the sign of the explicit
CP-violating term in the equation of motion; these lead to a
Chern-Simons condensate of unique sign across the universe.

We will give a few examples of numerical simulations, making no attempt to cover the possible range of initial conditions and models.  The examples are chosen to illustrate strong resonance and hence strong amplification of CP violation; in effect, this means that EW gauge potentials become of order the vacuum Higgs VEV $v$ and the Chern-Simons density grows to order $m^3/8\pi^2$.  In strong resonance these final values are essentially independent of the initial values; only the time needed to reach the final values changes with initial conditions.  With no resonance there is little or no amplification, and (in our scenarios) far too few baryons would be produced.  We are in no position to say what ``final'' values of the Chern-Simons condensate are needed to produce today's B+L values, since our understanding of the conversion of Chern-Simons number to B+L in this very non-equilibrium process is still quite primitive. 
 
Our general conclusions are that while there are still many unknowns, involving both
parameters of CP-violating physics and unsolved dynamics, we know of
nothing which would rule out generation of the observed number of baryons
in the universe today in this EW preheating scenario.

Other points of view \cite{kls,kt,ggks,rc} have been expressed concerning the mechanisms at work during EW preheating, including the invocation of approximate thermalization of long-wavelength gauge boson modes at an effective temperature much larger than the eventual reheat temperature, so that EW symmetry is effectively restored, and more or less genuine thermalization of shorter wavelengths.  Without further numerical simulations it is impossible for us to say how such ideas compare to what we suggest here, which is based on ultimate conversion of a spatially-homogeneous Chern-Simons condensate \cite{alexmike} to a condensate more resembling sphalerons.

There is one numerical lattice simulation \cite{ksc} of the full d=3+1 Higgs-gauge system which does not show any secular growth of Chern-Simons number, which instead decays to near zero.  For technical reasons these authors did not include CP-violating terms in the gauge equations of motion.  As we show here, such terms are important to establish a long-term Chern-Simons condensate, which may undergo sphalerization and conversion to baryons.

\section{ \label{spon} Spontaneous baryogenesis at preheating} 

Biasing the baryon asymmetry through an effective chemical potential can be
achieved in a model with two Higgs doublets in what is known as {\it
spontaneous baryogenesis}.\footnote{ For numerical simulations of
baryogenesis in two-Higgs models see \cite{Grigoriev:1992nv}.}  Cohen,
Kaplan, and Nelson~\cite{ckn,ckn1} proposed that the effective T-reversal
asymmetry may come from a time dependence in the solution for the Higgs
field.  Their scenario used the variation of the Higgs field inside a wall
of a bubble formed in a first-order phase transition.  A similar effect can
occur at preheating uniformly in space, on the horizon scales.  We will
discuss this scenario on an example of a two-doublet model.

The Higgs potential in a general model with two doublets, $H_1$ and $H_2$,
has a form

\begin{eqnarray}
V(H_1,H_2) & = & \lambda_1 (H_1^\dag H_1 -v_1^2)^2 \nonumber \\ & + &
\lambda_2 (H_2^\dag H_2 -v_2^2)^2 \nonumber \\ 
& + & \lambda_3 [(H_1^\dag H_1 -v_1^2)+ (H_2^\dag H_2 -v_2^2) ]^2 
\nonumber \\ 
& + & \lambda_4 [(H_1^\dag H_1)(H_2^\dag H_2)-
(H_1^\dag H_2)(H_2^\dag H_1)] \nonumber \\
& + & \lambda_5 [{\rm Re}(H_1^\dag H_2)-v_1 v_2 \cos \xi]^2 \nonumber \\
& + & \lambda_6 [{\rm Im}(H_1^\dag H_2)-v_1 v_2 \sin \xi]^2
\label{2higgs} 
\end{eqnarray} 
At finite temperature, all $\lambda_k$ and $v_i$ receive thermal
corrections and depend on the temperature.

During preheating the Higgs fields move along some classical trajectory 
\begin{equation}
H_i= \rho_i(t) e^{i \theta_i (t)}
\label{solution}
\end{equation}
that satisfies the equations of motion
\begin{eqnarray}
\ddot{\theta_i} + 3 h \dot{\theta_i} + \frac{\dot{\rho_i}}{\rho_i}+
\rho_i^{-1} \frac{\partial V}{\partial \theta_i} & = & 0, \nonumber \\
\ddot{\rho_i} + 3 h \dot{\rho_i} -\dot{\theta_i}^2{\rho_i}+ 
\frac{\partial V}{\partial \rho_i} & = & 0, 
\label{rho_theta_eqns}
\end{eqnarray}
where $h$ is the Hubble constant. The term $(\partial V/\partial \theta_i)$
is non-vanishing as long as there is a mixing between the Higgs bosons;
this term is periodic in $\theta_i$.

In the course of reheating, fermions are created and a thermal equilibrium
is achieved at some temperature $T_R<100$GeV, low enough to prevent any
sphaleron transitions.  The Higgs fields change from their zero-temperature  
values at the end of inflation to some temperature-dependent VEV: 
\begin{eqnarray}
{\rm at} \  T & = & 0,\ \  \rho_i=v_i; \\  {\rm at} \ T & =& T_{_R},
\ \ \rho_i=v_i(T_{_R}).
\end{eqnarray}
At the same time, the phase $\theta$ also changes: 
\begin{eqnarray}
\theta(0) & \equiv & \theta_1(0)-\theta_2(0)  = \xi, \\
\theta(T_{_R}) & \equiv & \theta_1(T_{_R})-\theta_2(T_{_R})=\xi(T_{_R}). 
\end{eqnarray}

The time derivative of $\theta$ serves as a chemical potential for the
baryon number~\cite{ckn1,comelli} because of an effective coupling 
%
\begin{equation}\label{missing}
(\partial_0 \theta) j^0_{_Y}
\end{equation}
that appears in the Lagrangian after the time-dependent phase is eliminated
from the Yukawa couplings.  Here  $j^\mu_{_Y}=m_t \bar{\psi} \gamma^\mu
\psi+...$ is the  fermionic current. 

If the gauge fields grow in resonance as described in Ref.~\cite{alexmike},
but the Higgs fields are out of resonance, the $\dot{\rho}/\rho$ term in
equation (\ref{rho_theta_eqns}) can be neglected.  The equation allows for a
slowly varying solution $\bar{\theta}(t)$ that interpolates between $
\theta=\xi$ and $\theta=\xi(T_{_R})$.  Since the Chern-Simon number
violating processes in the gauge sector are very rapid on the time scale of
thermalization, the fermions produced during reheating will have time to
equilibrate to the minimum of free energy when a thermal distribution is
ultimately achieved.  The effective chemical potential $\mu_{_B}$ is
proportional to $\dot{\theta}$, and the equilibrium value of baryon
asymmetry is  
\begin{equation}
\label{nB}
n_{_B} \sim \langle \dot{\theta} \rangle 
\left ( \frac{y_t v(T_{_R})}{T_{_R}}\right )^2 
T_{_R}^2 \sim \frac{\xi}{t_{_R}} 
\left ( \frac{y_t v(T_{_R})}{T_{_R}}\right )^2 
T_{_R}^2 \sim 10^{-10} \ T_{_R}^3 \ \left (\frac{10^{-5} t_{_H}}{t_{_R}}
\right) ,
\end{equation}
where $t_{_R}$ is the time of reheating and $t_{_H}$ is the Hubble time at
the electroweak scale, $y_t$ is the top Yukawa coupling, and $v(T)$ is the
Higgs VEV. For $T_{_R} \le 70$GeV, there is no wash-out of the
baryon number by thermal sphalerons.  The required baryon asymmetry,
$10^{-10}$, can be achieved in this scenario if the reheat time is $10^{-5}$
the Hubble time.  The reheat time is usually much shorter than the Hubble
time, and the requisite ratio $10^{-5}$ can be achieved in a realistic
model.

The difference with the scenario proposed by Cohen, Kaplan and
Nelson~\cite{ckn1} is that in our case CP violation occurs
homogeneously in space, as opposed to in a  bubble wall. In addition,
the final prediction for baryon asymmetry in the CKN scenario was very far
from the equilibrium value (\ref{nB}) because the sphaleron rate was slow
on the time scales associated with the growth of the bubble.  In our case,
$\theta$ changes slowly in time while the baryon number non-conservation
is rapid.  This allows a slow adiabatic adjustment of the baryon number to
that which minimizes the free energy.

\section{CP-Violating Operators and Coupling to the EW Gauge Fields}

In this section we describe some models of parametric-resonance-enhanced CP
and B+L violation which are (at least immediately following inflation)
spatially-homogeneous over each Hubble volume, but not fully-correlated
over super-horizon regions.  These models evade, in various ways, the usual
problem that B+L generation in any one Hubble volume is completely
uncorrelated to neighboring Hubble volumes.  In these models there is one
or more feature which is totally-correlated (because of inflation),
typically the inflaton VEV or the time-dependent part of Higgs VEVs as
driven by coupling to the inflaton.  But other features, such as initial
values of other fields, may be totally uncorrelated from one Hubble volume
to the next.  Unlike the earlier study \cite{alexmike} explicit CP
violation is built in to the gauge-field equations of motion rather than
just into initial values; this CP bias can overcome randomness due to
initial values and to stochastic behavior of the solutions to the equations
of motion.
 
The CP-violating terms come from various higher-dimensional operators which
can be generated from strong CP violation or from multi-Higgs models
showing spontaneous CP violation \cite{lw}. The coupling strengths of such
CP-violating operators need not be small, since in the early universe
strong CP violation can be very much out of equilibrium. Generally the
higher-dimension operators come from fermion loops; in the kinematic
situation of the present paper, there is no way for a purely bosonic system
of gauge fields, inflaton, and Higgs to lead to CP-violating terms in the
gauge equations of motion.  We will only consider $local$ CP-violating
terms, which would be appropriate at the beginning of pre-heating when the
universe is very cold; in principle, modifications of locality can occur
when the time dependence of various quantities begins to probe the fermion
loops carrying the CP violation.  We will not consider that case here.
 
There are two generic CP-violating operators which we will consider.  The
first is of the form
\begin{equation} 
\label{cpop1} 
\kappa F {\rm\,Tr\,} G_{\mu\nu}\tilde{G}^{\mu\nu}
\label{kappacoupling}
\end{equation} 
where $\tilde{G}$ is the dual field strength for the EW $SU(2)$
group\footnote{We ignore hypercharge couplings in this paper.  A factor of
the gauge coupling $g$ is included in our definition of EW potential and
field strength.}, $F(x)$ is a (possibly composite) gauge-singlet field and
$\kappa$ a coupling strength with dimensions of mass to some negative power
(depending on $F$).  As usual, integration by parts of equation
(\ref{cpop1}) shows that the action can contribute to the gauge-field
equations of motion only if $F$ has a non-trivial dependence on $x$; we
will seek for this dependence in out-of-equilibrium oscillations of the
inflaton or other fields.  We note that coupling of the form
(\ref{kappacoupling}) as a source for
topological number density has been considered in a model with a
cosmological pseudoscalar field coupled to hypercharge~\cite{brustein}.

Another possibility is: 
\begin{equation} 
\label{cpop2} 
\kappa' F{\rm\,Tr\,} G_{\mu\nu}G_{\nu\alpha}\tilde{G}_{\alpha\mu}. 
\end{equation}  
In this case, it is not required that there be any extra field $F(x)$,
since the operator in equation (\ref{cpop2}) is not a total divergence.
Nonetheless we include it, since there are terms in the gauge-field part of
(\ref{cpop2}) which are total divergences and which could be important.
 
When the action constructed from operators of the type (\ref{cpop1},
\ref{cpop2}) is added to the EW gauge action and the Higgs equation of
motion is added, the dynamics studied in \cite{alexmike} are modified.  We
discuss the new equations of motion following some brief remarks on the
physics behind the operators in (\ref{cpop1}).  As for the higher-dimension
operator in equation (\ref{cpop2}), it is to be expected at some level
whenever the lower-dimension operator in equation (\ref{cpop1}) appears.
 
\section{CP-Violating Physics} 
 
The CP-violating physics we are concerned with may come from
out-of-equilibrium strong CP violation, from CP violation in the Higgs
sector (with two or more Higgs fields), or from other causes.  (Note that
CKM phase effects are not strong enough to drive CP violation in the
Standard Model.)  We will discuss the first two explicitly.

\subsection{   A model with strong CP violation}

Consider the CP-violating operator of equation (\ref{cpop1}), which has a
typical axionic form, although we do not associate the field $F$ with an
axion.  We assume that previous physics associated with earlier times has
left a universe which has substantial strong CP violation in the QCD
sector.  As a prototypical example, consider the $\eta'$ field, which is
coupled to the QCD quark anomaly and to the gluonic topological charge
density.  This coupling to quarks is of the usual form
\begin{equation} 
\label{quark} 
M_q\bar{q}\exp [\frac{i\gamma_5\eta'}{F_{\eta'}}]q 
\label{etacoupling} 
\end{equation} 
(with suitable but irrelevant normalization of $F_{\eta'}$), showing that
$\eta'/F_{\eta'}$ is an angle with period $2\pi$.  The potential energy of
the $\eta'$ field comes from (\ref{quark}) and from coupling to gluons; the
latter must reflect the Witten-Veneziano \cite{witven} relation and other
requirements related to the $\theta$-angle dependence of QCD.  The result
is the standard form:
\begin{equation} 
\label{potential} 
V(\eta' )={\rm min}_j\;\varepsilon_{\rm QCD}\cos [\frac{1}{N_c}(\theta + 2\pi j+\eta'/F_{\eta'})] 
\end{equation} 
where $j=0,1,2$, $N_c=3$ is the number of colors, $\theta$ is the usual QCD
vacuum angle, $\varepsilon_{\rm QCD}$ is the QCD gluonic vacuum energy density
($\varepsilon_{\rm QCD}\simeq g_s^2\langle {\rm\,Tr\,} G_s^2\rangle $ in terms of the strong
coupling $g_s$ and field strength $G_s$; note that $\varepsilon_{\rm QCD}<0$ with our conventions using antihermitean gauge fields).  To be
specific, let us suppose that the vacuum angle is zero and that
$\eta'/F_{\eta'}$ is small enough so that we need consider only the $j=0$
sector explicitly.  Now suppose that through the operation of some
early-universe physics the field $\eta'$ deviates substantially from its
equilibrium value of zero.  Furthermore, this deviation, because of
inflation, is (roughly) the same across the entire universe.  Just as the
inflaton does, the $\eta'$ field will begin to roll toward its equilibrium
value (at which point strong CP violation is absent or small) on a QCD time
scale $\sim GeV^{-1}$, a time scale long compared to all other time scales
in the problem.  The $\eta'$ field couples to the EW fields $G_{\mu\nu}$
through, {\it e.g.}, quark loops, yielding an effective coupling
\begin{equation} 
\label{etaEW} 
\kappa_{\rm QCD}\eta'\frac{1}{32\pi^2}{\rm\,Tr\,} G_{\mu\nu}\tilde{G}^{\mu\nu}. 
\end{equation} 
It is straightforward to check that $\kappa_{\rm QCD}^{-1}$ is of order a QCD
mass $M$ of 1 GeV or so.  The coupling to the EW potential $\phi$ requires
an integration by parts, yielding a term in the action $\sim
\kappa_{\rm QCD}\dot{\eta'}\phi^3$, where $\phi$ is a constituent of the EW
gauge potential (see equation (\ref{ansatz}) below).
 
It might also happen that the field $F$ coupled to the topological charge
density oscillates at the inflaton rate, because of Higgs couplings to the
inflaton.  For example, with the notation $H$ for an electroweak Higgs
field doublet, some unspecified physics may lead to an action of the type
of equation (\ref{cpop1})
\begin{equation} 
\label{CP1} 
\kappa \frac{g^2}{16\pi^2}\int d^4x H^{\dag}H{\rm\,Tr\,}G_{\mu\nu}\tilde{G}^{\mu\nu} 
\end{equation} 
where the field $F=H^{\dag}H$ oscillates at inflaton rate scales because
the Higgs field is coupled to the inflaton.

\subsection{  CP violation in a multi-Higgs sector}

Since we are in no position to be very specific about CP violation in the
early universe, we will illustrate the concept by using a model
(T. D. Lee in \cite{lw}) which cannot account for all CP violation observed in the $K-\bar{K}$ system, but which can nevertheless represent an additional source of CP violation in the early universe.  In this model there are
two Higgs doublets $H_1,H_2$, and the potential is chosen so that even
though all couplings are real, the $H_2$ VEV is complex and of the form
\begin{equation} 
\label{h2} 
\langle H_2 \rangle = \left( \begin{array}{c}   0 \\ 
 v_2e^{i\theta} 
\end{array} \right) 
\end{equation} 
with real positive $v_2$. 
The phase angle $\theta$ arises from terms in the Higgs potential of the form 
\begin{equation} 
\label{cppot} 
V=H^{\dag}_1H_2(DH^{\dag}_1H_2+EH^{\dag}_1H_1)+H.c.+\dots 
\end{equation} 
which can be rewritten to show terms involving $\cos \theta$: 
\begin{equation} 
\label{thetapot} 
V=2Dv_1^2v_2^2(\cos \theta + \frac{Ev_1}{4Dv_2})^2+\dots  
\end{equation} 
 
Recall that the essence of EW-scale pre-heating is coupling of the inflaton
to the EW Higgs fields.  In this case, we invoke (for no deep physics
reasons) a coupling of the form
\begin{equation} 
\label{infhiggs} 
G\sigma^2H^{\dag}_1H_2+ H.c.\sim v_1v_2\sigma^2\cos \theta, 
\end{equation} 
assuming for simplicity  that the coupling $G$ is real.  Once
pre-heating sets in and the inflaton $\sigma(t)$ begins oscillating, the
angle $\theta$ also begins to oscillate.
 
The CP violation is coupled through $SU(2)$ fermions to the EW gauge
fields, and one readily checks that the lowest-order fermion loop graph
gives rise to a coupling of the type in equation (\ref{cpop1}) which is
proportional\footnote{The Higgs fields change left fermions to right, so
both $H_1$ and $H_2$ must act.} to $v_1v_2\sin \theta $.  The oscillations
of $\theta$ then give rise to a non-trivial coupling to the EW gauge field
$\phi$.
 
 \section{\label{cpeqns} CP-Violating Equations of Motion} 
 
There are several possibilities for CP-violating terms in the equations of
motion, depending, for example, on whether the field $F(x)$ in equation
(\ref{cpop1}) oscillates at the inflaton oscillation frequency, or at a
much lower frequency.  Before writing down these terms we set the notation
by giving the Standard Model action for the gauge and Higgs fields,
including a coupling of the Higgs field to the inflaton field $\sigma$ (but
not writing other inflaton terms):
\begin{equation} 
\label{smaction} 
S=\int d^4x\left[
\frac{1}{2g^2}{\rm\,Tr\,}G_{\mu\nu}G^{\mu\nu}+D_{\mu}H^{\dag}D^{\mu}H-\lambda
(H^{\dag}H-\frac{1}{2}v^2)^2-G^2\sigma^2H^{\dag}H\right]
\end{equation}  
We use, as before \cite{alexmike}, the following {\it ansatz} for the EW
gauge potential, which is $g$ times the canonical potential and is written
in antihermitean matrix form (so that $D_{\mu}=\partial_{\mu}+A_{\mu}$):
\begin{equation} 
\label{ansatz} 
A_0=0;\;\;A_j=(\frac{\tau_j}{2i})\phi (t), 
\end{equation}  
with corresponding electric and magnetic fields: 
\begin{equation} 
\label{EB} 
E_j\equiv G_{0j}=(\frac{\tau_j}{2i})\dot{\phi};\;\;B_j\equiv
\frac{1}{2}\epsilon_{jkl}G_{kl}=(\frac{\tau_j}{2i})\phi^2.
\end{equation} 
For the Higgs field $H$ we use the Standard Model form\footnote{The form of
equation (\ref{higgs}) does not contain Goldstone modes, which contain
vital topological information during actual creation of baryons; we will
discuss these effects in a later section.}
\begin{equation} 
\label{higgs} 
\langle H \rangle = \left( \begin{array}{c}   0 \\ 
 v(t)/\sqrt 2 
\end{array} \right) 
\end{equation}

Evidently the topological charge density goes like $\dot{\phi}\phi^2$,
which is, as it must be, a total time derivative.  The topological charge
density is the time derivative of the Chern-Simons density $\ncsmall$:
\begin{equation} 
\label{csno} 
\ncsmall =\frac{\phi^3}{8\pi^2}. 
\end{equation}

The gauge and Higgs equations will be written in non-dimensional form,
where the time variable is replaced by $mt$, the gauge field $\phi$ is
replaced by $\phi /m$, and the Higgs field $v$ is replaced by $v/v_0$.
Here $m=v_0g/2$ is the vacuum W-boson mass in terms of the standard Higgs
VEV $v_0$.  The resulting\footnote{In these equations we ignore Hubble damping, which is miniscule, and damping by decay into fermions, which may have some importance at long time scales where other effects we ignore could also be important.} equations are:
\begin{equation}
\label{newmotion}
\ddot{\phi}+2\phi^3+v^2\phi+[{\rm CP-violating\;term}]=0 
\end{equation}
and
\begin{equation}
\label{higgsmotion}
\ddot{v}+\frac{3}{4}\phi^2v+(\frac{M_H^2}{2m^2})
v(v^2-1)+(\frac{G^2\sigma^2}{m^2})v=0.
\end{equation}
where $M_H=v_0\sqrt {2\lambda} $ is the Higgs mass.  An important property of equation (\ref{higgsmotion}) is that if large (order one) values of $\phi$ are reached by resonant amplification, the Higgs potential is strongly modified and $v$ can approach the symmetry-restoring value of zero.  

There are, as
discussed above, several possibilities for the CP-violating term.  In these
equations we will assume that the inflaton is oscillating sinusoidally at a
physical frequency $\omega$, corresponding to a dimensionless frequency
$r/2$:
\begin{equation}
\label{sigma}
\sigma =\sigma_0\cos (\frac{rt}{2});\;\;\omega = \frac{rm}{2}.
\end{equation}
Previously instead of using the Higgs equation of motion
(\ref{higgsmotion}) the Higgs potential was set to zero, on the grounds
that inflation stops, preheating starts, and the EW sector begins to
undergo spontaneous symmetry breaking when the inflaton field reaches the
critical value $\sigma_c$:
\begin{equation}
\label{sigmac}
\sigma_c=\frac{M_H}{G\sqrt 2}.
\end{equation}
This led to
\begin{equation}
\label{oldv}
v^2=1+2\epsilon +2\epsilon \cos rt;\;\; \epsilon \equiv
\frac{-\sigma_0^2}{4\sigma_c^2}.
\end{equation}
Under the assumption that $\epsilon$ is small,\footnote{It is useful for
approximate analysis to take the parameter $\epsilon$ to be small, but it
can be of order unity in actuality.} so that it can be ignored compared to
unity in the constant term in (\ref{oldv}), the gauge equation of motion
became \cite{alexmike}:
\begin{equation} 
\label{oldmotion} 
\ddot{\phi}+2\phi^3+(1+\epsilon \cos rt)\phi=0. 
\end{equation}

Note that there is no CP violation in equation (\ref{oldmotion}), since for
any solution $\phi (t)$ there is another solution $-\phi (t)$.  CP
violation can only come from initial values which favor one sign or the
other of the CP-odd field $\phi$.  The modifications we address in the next
subsection add terms even in $\phi$ to the equations of motion, which
therefore will have a built-in CP bias.  And when we come to numerical
analysis in Sec. 5, we will use equations (\ref{newmotion},
\ref{higgsmotion}) instead of the simplified form $1+\epsilon \cos rt$ in
equation (\ref{oldmotion}) above.

\subsection{  \label{oscil} CP violation at the inflaton oscillation rate}

Before studying the complications of the combined Higgs-gauge equations
(\ref{newmotion}, \ref{higgsmotion}) which can really only be handled
numerically, we will look at adding CP-violating terms to the simplified
gauge dynamics of equation (\ref{oldmotion}).  This allows a certain amount
of approximate analysis, similar to that used in the Mathieu equation, to
be done.
 
Using the Higgs dependence on time as shown in (\ref{oldv}) and
integrating by parts in the action of equation (\ref{CP1}) gives rise to an
action of the form $\int \phi^3\sin \omega t$.  We add the appropriate
contribution to equation (\ref{oldmotion}) leading to the modified (and
explicitly CP-violating) equation:
\begin{equation} 
\label{motion} 
\ddot{\phi}+2\phi^3+\delta \phi^2\sin rt +(1+\epsilon \cos rt)\phi=0. 
\end{equation} 
In this equation the parameter $\delta$ is proportional to $\kappa \epsilon
\omega v_0^2$. Note that if $\phi(t)$ is the solution to (\ref{motion}), then
$-\phi(t)$ is the solution to this equation when $\delta$ is changed in
sign.

Without the $\delta \phi^2$ term the equation of motion is essentially a
Lam\'e equation, as discussed in \cite{alexmike}.  But with this term added
we know of no way of reducing the equation of motion to explicitly-soluble
form.  The major feature of the modified equation can be found by a
standard Mathieu-like analysis, assuming that $\phi$ has the form
\begin{equation} 
\label{mathieu} 
\phi (t)= a(t)\cos (rt/2) + b(t) \sin (rt/2) +c(t) 
\end{equation} 
and assuming that $a,b,c$ vary slowly on the time scale of oscillations.
The new feature here is the presence of the secular term $c$, which is
unnecessary when the $\delta \phi^2$ term is missing in (\ref{motion}).  By
ignoring all terms oscillating faster than the half-frequency $r/2$ one
then finds:
\begin{eqnarray} 
\label{abc} 
0 & = & r\dot{a}
+b[\frac{r^2}{4}+\frac{\epsilon}{2}-1-\frac{3}{2}(a^2+b^2)]-\delta ac; \\
\nonumber 0 & = &
-r\dot{b}+a[\frac{r^2}{4}-\frac{\epsilon}{2}-1-\frac{3}{2}(a^2+b^2)]-\delta
bc; \\ \nonumber c & = & \frac{-\delta ab}{1+3(a^2+b^2)}.
\end{eqnarray} 
Note that since both $\delta$ and $c$ are proportional to the
parametric-resonance parameter $\epsilon$, the new (last terms on the
right-hand side of (\ref{abc})) contributions are of order $\epsilon^2$.
The formal analysis is done in powers of $\epsilon$, which we therefore
assume to be small.  Then in (\ref{abc}) the equations for $a$ and $b$ are,
to order $\epsilon$, the same as without the $\delta$ term, and the secular
term $c$ is driven by the unperturbed values of $a,b$.  These show standard
Mathieu behavior of the form $\exp \mu t$ with
\begin{equation} 
\label{mu} 
\mu=\pm \frac{1}{2r}[\epsilon^2 -(r^2-4)]^{1/2} 
\end{equation} 
(where we ignore temporarily non-linearities in the equations).  Since
$\epsilon$ is small, the value of $r$ must be nearly two for $\mu$ to be
real and resonant amplification to take place, which is to say that $\mu$
scales linearly with $\epsilon$.
 
In studying the original equation with no $\delta$ term but retaining the
cubic non-linearity \cite{alexmike}, it was shown that the quantity
$a^2+b^2$ obeyed a certain equation which revealed the conditions under
which growth rather than damping (negative $\mu$) occurred.  With the
$\delta$ term added, this equation is:
\begin{equation} 
\label{ampgrowth} 
\frac{d}{dt}(a^2+b^2)=\frac{-2\epsilon ab}{r}-\frac{2\delta^2
ab(a^2-b^2)}{1+3(a^2+b^2)}.
\end{equation} 
The second term on the right of (\ref{ampgrowth}) is of higher order and
can be neglected, and then the conclusion is the same as before: Growth is
only possibly if $ab$ is positive (for negative $\epsilon$); otherwise there is damping.  (When $a,b$
are small, their growth leads to growth of the secular term $c$, at twice
the rate of growth of $a$ or $b$; see (\ref{abc}).)

Writing 
\begin{equation} 
\label{abdef} 
a=A\cos \Psi, b=-A\sin \Psi 
\end{equation} 
with positive amplitude $A$,  Ref. \cite{alexmike} showed from (\ref{ampgrowth})
that\begin{equation}
\label{Apsi} 
A=A_0\exp \int_0^tdt'\frac{\epsilon}{2r}\sin 2\Psi(t') 
\end{equation} 
with an accompanying equation for the angle $\Psi$ showing that it varied
on the $\epsilon$ rate scale, that is, slowly.  Eventually $\Psi$ becomes
large enough so that the product $ab$ changes sign and growth turns into
damping, or vice versa.

What is happening to the Chern-Simons number in this model?  The
Chern-Simons number density is just $\phi^3/8\pi^2$, and in the case where
the CP-violating term proportional to $\delta$ is absent, as is the secular
term $c$, this quantity is purely oscillatory and has no appreciable
long-term average or preferred sign.  But things are different with
$\delta\neq 0$; the secular average of the Chern-Simons density is:
\begin{equation} 
\label{CSav} 
\langle \frac{\phi^3}{8\pi^2}\rangle = \frac{3}{16\pi^2}cA^2 
\end{equation} 
Note that the sign of the Chern-Simons density is controlled by the sign of
the secular term $c$.  Under the usual assumption that the inflaton (and
therefore the Higgs) field is correlated across the entire universe,
because of inflation, the sign of $\delta$ is also correlated across the
universe and by (\ref{abc}) the sign of $c$ depends on the sign of the
product $ab$, which may differ in each Hubble volume.  But by equation
(\ref{ampgrowth}) growth only occurs when this product is positive (since $\epsilon$ is negative), even
though $a,b$ may separately be random in each Hubble volume.  Of course,
only those Hubble volumes where growth actually takes place have
appreciable Chern-Simons density and in each of these volumes the sign of
the Chern-Simons density is the same. There are Hubble volumes with
Chern-Simons density of the opposite sign, but in them there is no growth
and this density is small.  The result is that at least for a while {\em the entire universe has
an average Chern-Simons density of fixed sign, the same in each Hubble
volume, even though the initial values of gauge potentials may be random in
each Hubble volume}. 

Numerical simulations discussed below show that on time scales of order $\epsilon^{-1}$ the secular average can change sign, since on such time scales the product $ab$ can also change sign.  This may lead to relatively small Chern-Simons condensate values even in strong resonance.

We cannot yet say how small or large this condensate value must be, in order to reproduce the observed value of B+L, or of the baryon-photon ratio. 
The smallness of the baryon-photon ratio must come, in our models, from a
combination of the smallness of CP-violating parameters (such as occur in
Higgs potentials), dynamical effects during pre-heating such as mentioned above, and sphaleron
washout after pre-heating.

 \subsection{  Equations of motion with strong CP violation} \label{4B}

Since the $\eta'$ field potential in equation (\ref{potential}) is
determined by QCD parameters it is rolling very slowly on EW time scales
and we can replace the time derivative of $\eta'$ by a term of order
$M\eta'$, where $M$ is a typical QCD mass, and ultimately (after rescaling
time and the EW potential $\phi$ as before) come to another CP-violating
$\phi$ equation of the form:
\begin{equation} 
\label{etamotion} 
\ddot{\phi}+2\phi^3+\delta' \phi^2+(1+\epsilon \cos rt)\phi=0. 
\end{equation} 
Here the parameter $\delta'$ is of order $M/m$, the ratio of QCD and EW
scales, or perhaps $10^{-2}$.
 
As before, we must add a secular term $c$ to the usual Mathieu {\it ansatz}
of equation (\ref{mathieu}).  Equations (\ref{abc}) are changed; the
equations of motion for $a,b$ are the same as with no CP violation (that
is, set $\delta=0$ in the first two equations of (\ref{abc})) while the
equation for $c$ becomes:
\begin{equation} 
\label{ceqn} 
c=\frac{-\delta'(a^2+b^2)}{2[1+3(a^2+b^2)]}. 
\end{equation} 
Provided that (as we have already assumed) $\delta'$ is roughly the same in
all Hubble volumes, so is the sign of $c$, and so the Chern-Simons numbers
in all Hubble volumes add with the same sign.

\subsection{  
Higher-derivative CP violation}

We briefly consider here the consequences of adding the
higher-derivative CP violation of equation (\ref{cpop2}).  It is
readily checked that this term yields actions $\sim \int
F(\dot{\phi}^3)$ and $\sim F\phi^4\dot{\phi}$; the latter is
integrated by parts to $\sim \dot{H}\phi^5$.  So to the original
equations of motion we must add terms of the form
\begin{equation} 
\label{ddot} 
\dot{F}\phi^4;\;\;-3(\dot{F}\dot{\phi}^2+2F\dot{\phi}\ddot{\phi}). 
\end{equation} 
These can, as discussed in connection with the numerical studies reported below, have
dramatic effects because of the appearance of derivatives of $\phi$.

\section{Numerical studies}

The range of parameter space is vast, and cannot be covered in depth here.
We give several examples illustrating most of the effects discussed above.
In these examples, we will use the following values of parameters unless
otherwise specified:
\begin{enumerate}
\item Gauge initial values:  $\phi(0)=0.01;\;\dot{\phi}(0)=0.01$
\item Higgs initial values:  $v(0)=0.5;\;\dot{v}(0)=0.5$
\item Epsilon parameter\footnote{We choose $\epsilon$ positive for the simulations; this makes no difference to the ultimate interpretation of the numbers.} $\epsilon=0.5$
\item $M_H^2/2m^2=1.5$ (corresponding to a Higgs mass of about 140 GeV)
\item Resonance parameter $r=2.2$
\end{enumerate}
Other parameters will be specified as needed. Note that for this standard set of parameters the Higgs field is far from equilibrium; below we give numerics for initial values starting rather close to equilibrium.

\subsection{ \label{sin} Sinusoidal CP violation with gauge backreaction}

Gauge backreaction is described by equations (\ref{newmotion},
\ref{higgsmotion}), to which we will add appropriate CP-violating terms.
One of the most important features of this gauge backreaction is that it
can help facilitate rapid Higgs transition through $v=0$, which removes the
sphaleron barrier and makes baryons from Chern-Simons number.  The
equations to be studied numerically here are (\ref{higgsmotion}) plus a
gauge equation with a sinusoidal CP-violating term:
\begin{equation}
\label{sinnum}
\ddot{\phi}+2\phi^3+v^2\phi+\delta \phi^2\sin rt=0. 
\end{equation}

\begin{figure}[!ht]
\centerline{\epsfxsize 5 truein \epsfbox{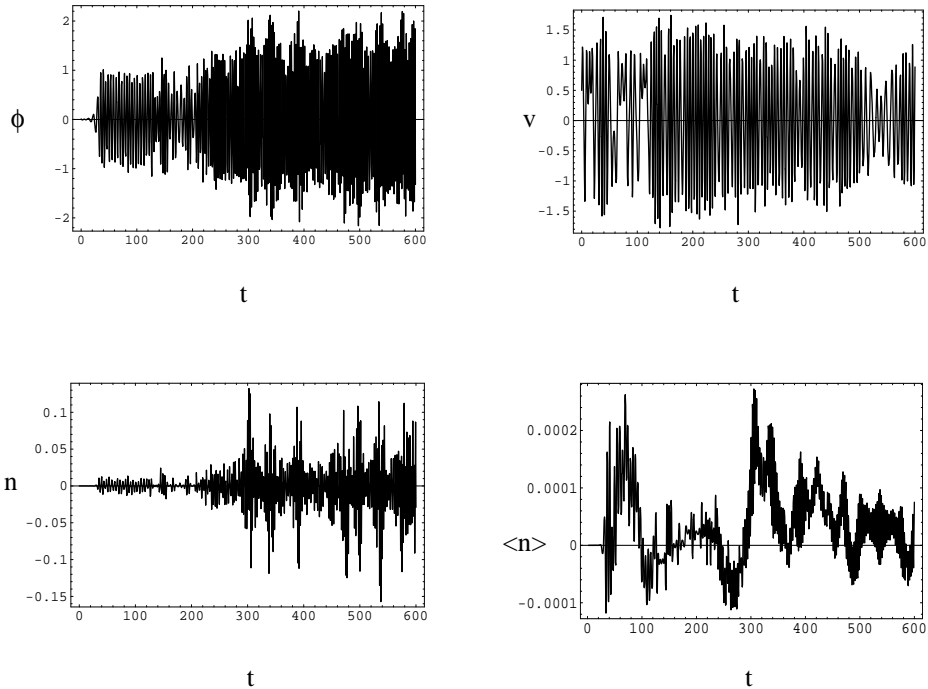}}
\caption[]{\small \label{cp1fig1}  These figures show the behavior of the dimensionless gauge potential $\phi$, Higgs VEV $v$, Chern-Simons density $n\equiv \ncsmall$, and the
running average of the Chern-Simons density $\langle n \rangle \equiv \ncsa$ for the conditions of Sec. \ref{sin}.} 
\end{figure}

The initial conditions are as given at the beginning of this Section,
except that $\dot{\phi}(0)=-0.01$; also, $\delta =0.51$.  This value of $\delta$ is not necessarily realistic, but in general CP-violating quantities scale linearly in $\delta$.  The change of
sign is done in order to get on the positive-$\epsilon$ growth curve
of the parametric-resonance instability of the simplified version of
the gauge equation of motion, given in equation (\ref{motion}).
Fig.~\ref{cp1fig1} shows the results for $\phi$, $v$, Chern-Simons
density $\ncsmall$, and 
the running average $\ncsa$ of the Chern-Simons density, defined as:
\begin{equation}
\label{secav}
\ncsa = \frac{1}{t}\int_0^tdt'\ncsmall(t').
\end{equation}
There is no particular physical significance to the running average, but it is more convenient to display than the time-integrated Chern-Simons number, which covers too large a range to be displayed legibly.  More to the point is a simple time integral of $\ncsmall$ such as emerges from the dynamics, as in equation (\ref{ncsavg}) below; this is 600 times the running average at the end of the time period simulated.

One notes that in the present model the running 
 average $\ncsa$ itself changes sign  from time to time on a long time scale, as would be
expected from the analysis of Sec.~\ref{oscil}. So this model is not a particularly efficient way of generating B+L even in strong resonance.  On the other hand, just how large a long-term Chern-Simons condensate needs to be to generate the observed B+L is not known, so this may or may not be a drawback.

\begin{figure}[!htb]
\centerline{\epsfxsize 5 truein\epsfbox{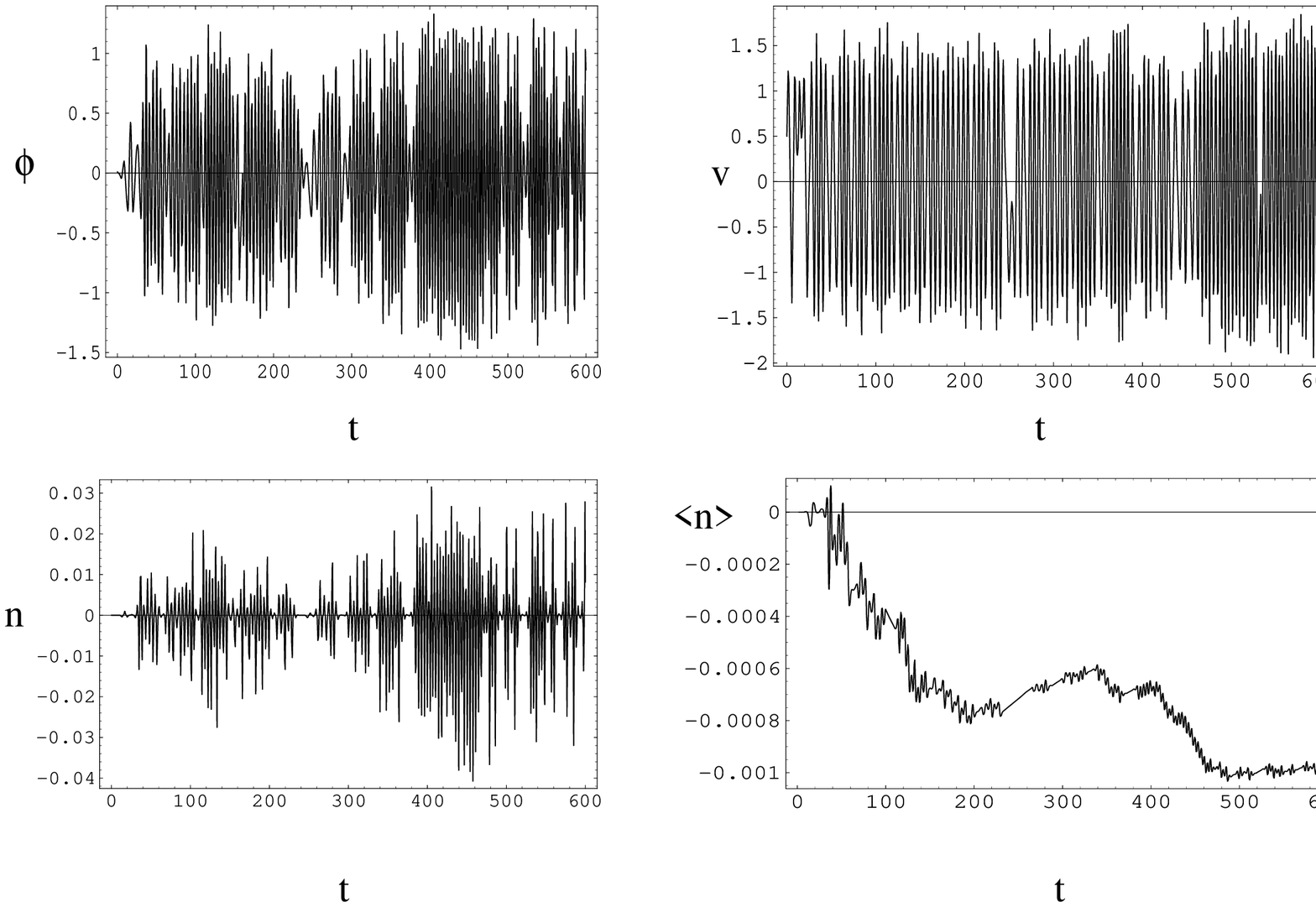}}
\caption[]{\small \label{cp1fig2}  These figures show the behavior of $\phi ,\;v ,\;\ncsmall, $ and $\ncsa$ for the conditions of Sec. \ref{const} with $v(0)=\dot{v}(0)$=0.5.}
\end{figure}

\begin{figure}[!ht]
\centerline{\epsfxsize 5 truein\epsfbox{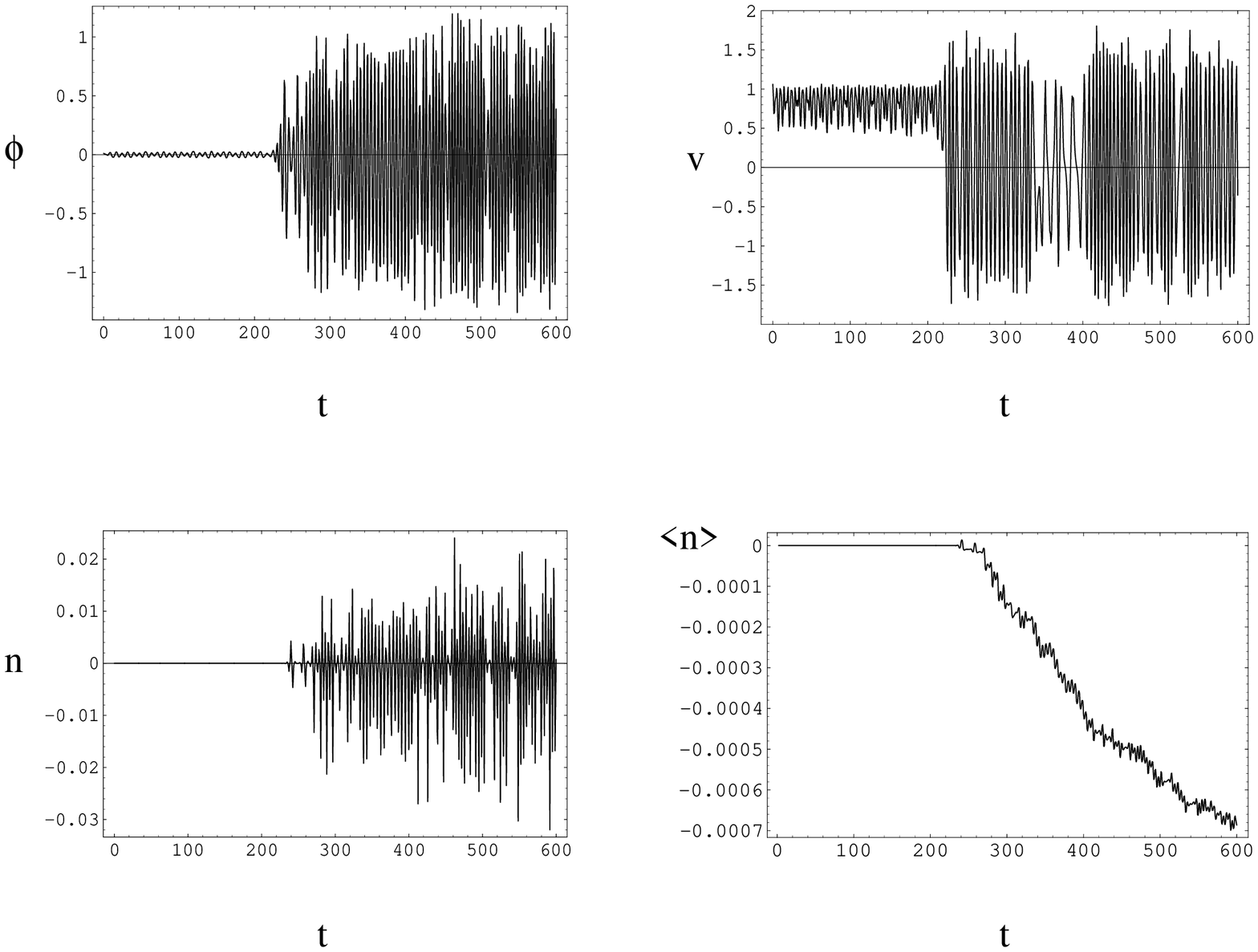}}
\caption[]{\small \label{cp1fig3} These figures show the behavior of $\phi,\;v,\;\ncsmall,\;$ and $\ncsa$ for the conditions of Sec. \ref{const}, with $v(0)$=1.06, $\dot{v}(0)$=0.1.}
\end{figure}

\subsection{  \label{const} Constant CP violation with gauge backreaction}

In the present example we modify equation (\ref{newmotion}) by adding a
quadratic CP-violating term with constant coefficient, as discussed in
Sec.~\ref{4B}: 
\begin{equation}
\label{newmotion1}
\ddot{\phi}+2\phi^3+v^2\phi+w\phi^2=0 
\end{equation}
We will take $w=0.5$ for the numerical example.  This is unrealistically
large, but it makes it easier to see what is happening.  Generally
speaking, the long-term average Chern-Simons density is proportional to $w$; for example, as in  the simplified analysis (with no gauge back-reaction) leading to equations~\eq{CSav},\eq{ceqn} which gives:
\be
\label{simplecp}
\ncsa = -{1\over{16\pi^2}}w\langle\phi^2\rangle
\ee
or $\ncsa\sim -0.0008$ for Figs.~\ref{cp1fig2} and \ref{cp1fig3}.

Fig.~\ref{cp1fig2} shows the time history of $\phi,\;v$, the Chern-Simons
density $\ncsmall$, and the running average $\ncsa$ for initial Higgs values which are far from equilibrium.  The integrated Chern-Simons number is of order 0.1
 for a CP-violating strength of order unity, and so might be of
order $10^{-4}$ for more realistic CP-violating amplitudes $w$ of order
$10^{-3}$.  This leaves room for inefficiencies in generating B+L from Chern-Simons number, washout of B+L, and other dynamical
effects to reduce the baryon number to its actual level of perhaps
$10^{-10}$.  And, of course, there is no reason to believe that the
numerical parameters we use here apply to the real early universe.
Experience with running many simulations of the type in Fig.~\ref{cp1fig2}
shows that the initial values of Higgs and gauge fields are not so
important, and that final values of the running Chern-Simons average for unit-strength CP violation range from
about $10^{-3}$ (somewhat smaller than the maximum expected, which is of order $1/8\pi^2$), to the non-resonant value of $\phi (0)^3/8\pi^2$ which is of order $10^{-8}$ for our initial conditions.  Of course, these values would be reduced further by the strength of the CP-violating term, which is not small in our simulations.

In the examples so far, the Higgs field has been far from equilibrium
initially and so it swings through the origin frequently.  The next
example, also for a constant CP-violating term, shows the importance
of gauge backreaction in causing the Higgs field to oscillate through
zero.  For this example we choose Higgs initial values
$v(0)=1.06,\;\dot{v}(0)=0.1$ with all other parameters as for
Fig.~\ref{cp1fig2}.  The results are shown in Fig.~\ref{cp1fig3}.
Because the Higgs field is near equilibrium, it stays for a while near
unity, and the gauge field appears to be trivially small.  Actually
$\phi$ is slowly growing and  finally becomes large around a time
of 200.  Fast growth of $\phi$ towards order unity values begins to
send the Higgs field into wild oscillation, accompanied by 
growth of the Chern-Simons average value.  We will see later that
having the Higgs field go through the origin may be important for
creating baryons from the Chern-Simons condensate.

\begin{figure}[!ht]
\centerline{\epsfxsize 5 truein\epsfbox{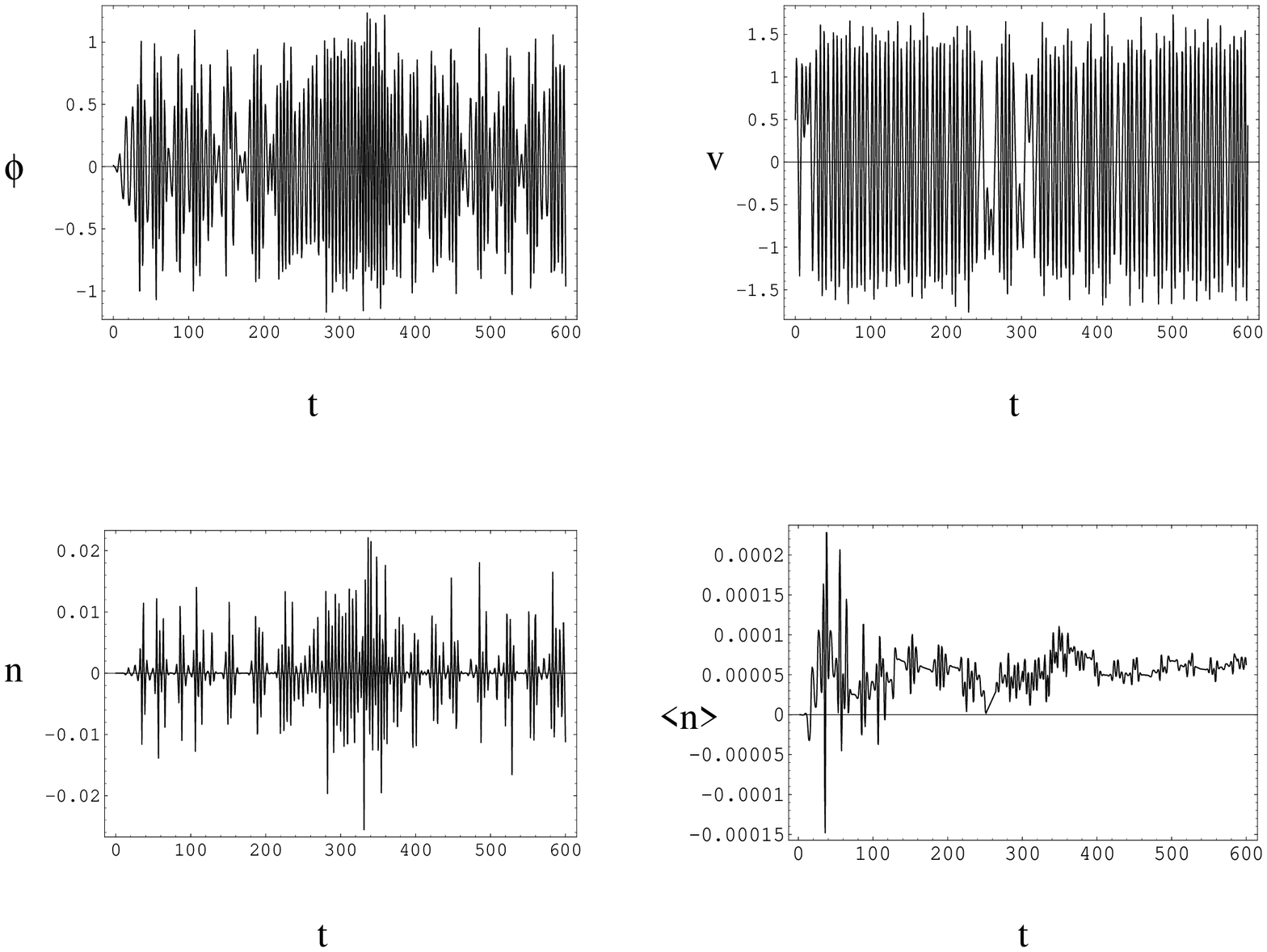}}
\caption[]{\small \label{cp1fig4} These figures show $\phi,\;v,\;\ncsmall,$ and $\ncsa$ for the conditions of Sec. \ref{higher}.}
\end{figure}

\subsection{ \label{higher} Higher-derivative CP violation}

Adding terms such as those in equation (\ref{ddot}) can and sometimes does lead to singularities in finite time because of the higher derivatives.  We show one case where the singularity does not develop before dimensionless time 600.  The equations to be solved numerically are the Higgs equation (\ref{higgsmotion}) and:
\begin{equation}
\label{ddoteqn}
\ddot{\phi}+2\phi^3+v^2\phi+w (\alpha \dot{\phi}^2+ \dot{\phi}\ddot{\phi})=0.
\end{equation}
The coefficients $w,\alpha $ are -0.2 and 0.1 respectively.  If $w$ gets even a little larger in magnitude (say, $w=-0.21$) a singularity seems to develop, although we have not explored this phenomenon in any detail.  Fig.~\ref{cp1fig4} shows the results.  The $\dot{\phi}^2$ term in (\ref{ddoteqn}) is always of one sign and leads to secular CP violation; if only the $\dot{\phi}\ddot{\phi}$ term were kept, the running-average CP violation slowly wanders from one sign to the other.

\section{From Chern-Simons Number to Baryons }

In this section we comment on some very difficult questions: How is
the homogeneous condensate of Chern-Simons number,
\footnote{Throughout this section we will make a distinction between total Chern-Simons
number $\ncs$ or Higgs winding number $\nw$ in a fixed volume $V$ of
space (or length $L$ in 1+1 dimensions) and their densities
$\ncsmall,\;n_{\rm wind}$, and similarly between the total energy $E$
in $V$ and its density $\varepsilon$.  Note that $E/V$ is a function
of $\ncsmall,\;n_{\rm wind}$ but not of $V$; nonetheless, it is
appropriate to discuss the dependence of $E$ on the total topological
numbers $\ncs,\;\nw$ at fixed $V$.} equivalent to a B+L condensate,
turned into actual baryons and leptons, which are localized states?
What fraction of the condensate is turned into baryons and leptons?
How does the winding number of the Higgs field enter in?  We are in no
position to give definitive answers; all these questions are still
under active investigation.

Let us begin with a comment concerning washout of baryons and leptons after reheating. 
The attractive feature of EW preheating is that the temperature after full
reheating is rather smaller than the EW cross-over temperature $T_c$,
so that baryons and leptons created during preheating are not completely washed out
(the sphaleron transition rate $\sim \exp (-M_S/T)$ is very small because
the sphaleron mass $M_S$ is large compared to the temperature).  We will
assume that this mechanism protects baryons, once produced, from washout during reheating.  Nonetheless, there can be washout during preheating, and we will have to discuss that.

Our discussion will invoke insights derived from numerical and analytical work on these issues in the 1+1-dimensional Abelian Higgs model; from simple arguments based on effective temperature; and on approximations to an  analysis based on modifications of earlier work on B+L violation in two-particle collisions (where B+L violation is still extremely small because of the poor overlap of states, even though there may be no tunneling barrier).

A real baryon is a spatially-localized state, so the transformation to baryons requires quasi-localized states resembling (but not necessarily identical to) sphalerons; these states involve not only the gauge fields but also Higgs fields, which carry a winding number $\nw$ of their own.  In the {\it ansatz} we have used so far the Higgs field has no  winding number.

Presumably a major influence on the dynamics of spatial localization is the instability of the gauge-Higgs equations to growth of spatial ripples, as alluded to in \cite{alexmike}.  We will not discuss this mechanism here, which is non-linear and time-dependent (so it may be amplified by parametric resonance). 

\subsection{Dependence of energy on topological quantum numbers}  

In the presence of a Chern-Simons 
condensate it is energetically favorable to shift the winding number
of the Higgs field $\nw$ in the direction of Chern-Simons number.  One way to see this is to note that the energy  $E=\int d^3x\varepsilon$ depends on both $\ncs$ and $\nw$.  In the homogeneous $ansatz$ used so far the energy, with terms in $\phi^2$ and $\phi^4$, grows monotonically with $\ncsmall$ since $\ncsmall \sim \phi^3$.  The Higgs winding number $\nw$ is zero (see equation (\ref{higgs})).  Now the energy  of a state with finite values of $\nw,\;\ncs$ is, under a large gauge transformation with Chern-Simons number $-\nw$, equivalent to one with zero Higgs winding:
\be\label{eminus1}
E(\ncs,\nw)=E(\ncs-\nw,\nw=0)
\ee 
By the above remarks the energy with $\nw=0$ has its minimum at $\ncs-\nw =0$.

More explicitly, once the system has become a condensate of more or less localized objects, these are described by a large gauge transformation at spatial infinity:
\be
\label{gauge}
A_{\mu}\rightarrow U\partial_{\mu}U^{-1},
\ee
\be
\label{higgs2}
H(x)\rightarrow \frac{1}{\sqrt 2}U\left( \begin{array}{c}   0 \\ 
 v_0 
\end{array} \right).
\ee  
The same gauge transformation $U$ appears in both the gauge potential and in the Higgs field in order that the Higgs contribution $\int |D\phi|^2$ to the action be finite in infinite volume.\footnote{The gauge transformation $U$ has the usual properties of large gauge transformations, including compactness on the sphere at infinity; a typical example is given in equation (\ref{bc}) below.}  Evidently these localized states have lower energy than a homogeneous condensate which, so to speak, fills in the volume between the localized states.  

The vacuum  states with $\ncs =\nw$ are minima on the energy landscape, so $E(\ncs ,\nw )$ is zero along the line $\ncs =\nw$.  Any transitions proceeding toward this line are energetically favorable.  For example, beginning from the homogeneous $ansatz$ state with large positive $\ncs$ but zero $\nw$, it is favorable to $decrease$ $\ncs$ and $increase$ $\nw$.   Of course, it is also energetically favorable to decrease $\ncs$ toward zero with no change in $\nw$, but then no baryons will be produced; B+L will rise and fall in lockstep with $\ncs$, according to the usual anomaly relation, yielding no baryons at the end.  So among the changes which decrease $\ncs$ there is a competition between those which increase $\nw$ and those which do not change $\nw$.

\begin{figure}[!t]
\centering
\hspace*{-5.5mm}
\leavevmode\epsfysize=8.5cm \epsfbox{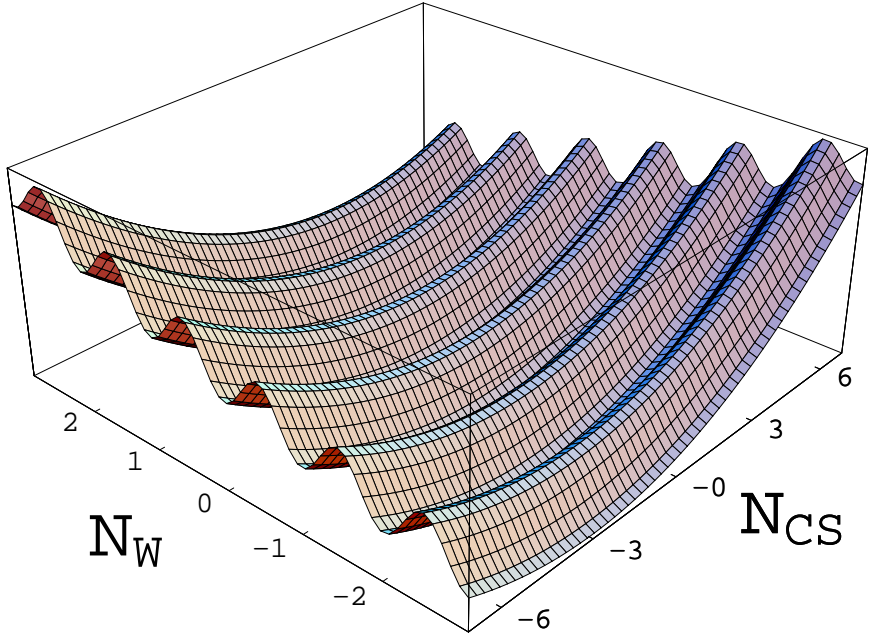}\\[3mm]
\caption[fig9]{\label{figNN} \small The static system energy $E$ as a function of the
Chern-Simons number {$\ncs$} and the winding number {$\nw$}.  Note that $E(\ncs ,\; \nw )$ is minimal along the vacuum line $\ncs =\nw$ and increases with $|\ncs -\nw |$. 
}
\end{figure}

\begin{figure}[!t]
\centering
\hspace*{-5.5mm}
\leavevmode\epsfysize=8.5cm \epsfbox{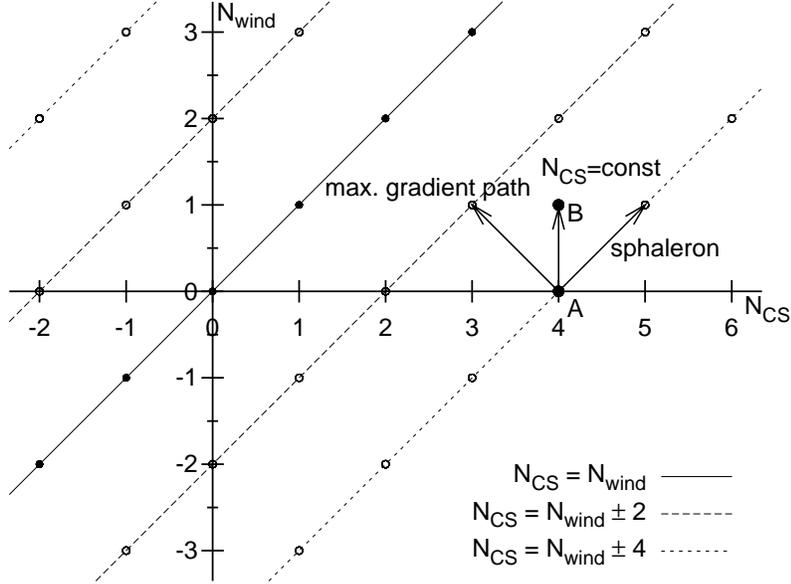}\\[3mm]
\caption[fig10]{\label{fromabove} \small Possible paths of topological
transitions are shown on the $\ncs$ -- $\nw$ diagram, viewed from above.  Path AB represents the simplest non-sphaleronic topological transition in the Higgs sector.  See the text for explanation of the notations.}
\end{figure}

\begin{figure}[!t]
\centering
\hspace*{-5.5mm}
\leavevmode\epsfysize=8.5cm \epsfbox{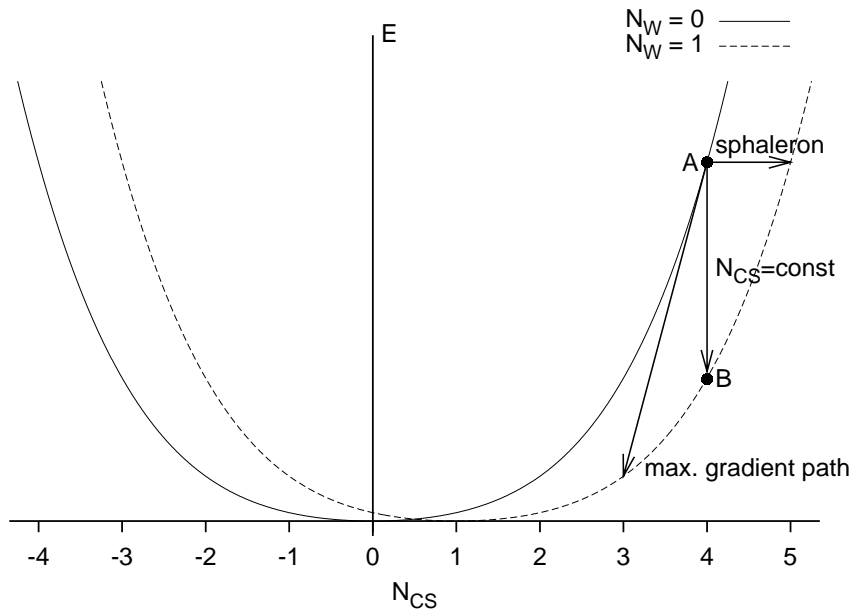}\\[3mm]
\caption[fig3]{\label{ncsplot} \small  A cross-section of the energy profile at constant $\nw$.  The lack of any barriers makes the downward evolution of $\ncs$ completely unsuppressed.  States A and B correspond to the same states in Fig. \ref{fromabove}.  At non-zero $\ncs$ the transitions which decrease $|\ncs -\nw |$ become energetically favorable; a simple case discussed in the text corresponds to the path AB with $\ncs$ held constant.
}
\end{figure}

\begin{figure}[!t]
\centering
\hspace*{-5.5mm}
\leavevmode\epsfysize=8.5cm \epsfbox{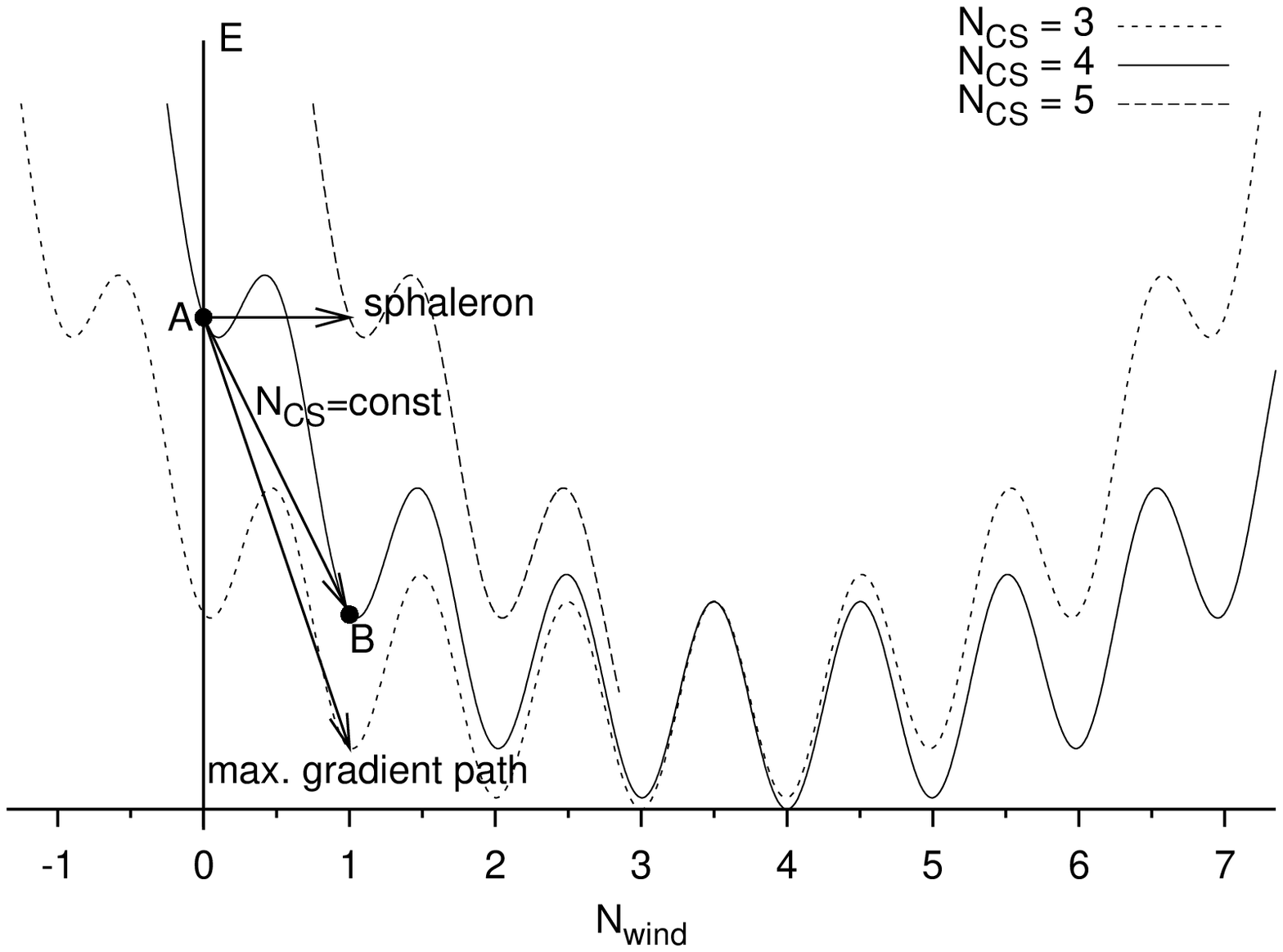}\\[3mm]
\caption[fig4]{\label{nwplot} \small Energy profile $E(\ncs,\nw)$ along the $\nw$
axis for several values of $\ncs$. Note that $\nw$ is integer for
$any$ field configuration; the curve profile at non-integer $\nw$
reflects the presence of sphaleron-like energy barriers separating the
states with $\Delta\nw = \pm 1$. These barriers disappear at sufficiently high $|\ncs -\nw |$.}
\end{figure}

We illustrate these statements in Figs. \ref{figNN}, \ref{fromabove}, \ref{ncsplot}, \ref{nwplot}.  Fig. \ref{figNN} is drawn assuming that the gauge boson mass (and {\it a posteriori} the sphaleron mass) are somewhere near their conventional values, so that there is a sphaleron barrier hindering topological charge change as indicated by the corrugations on the figure.  However, as we have seen in the numerical studies, it is possible for this barrier to vanish as the Higgs VEV passes through zero.  Fig. \ref{fromabove} is a view from above of the energy profile plot of Fig. \ref{figNN}.  In Fig. \ref{fromabove} the filled circles along the $\ncs=\nw$ line indicate states of zero energy (with $\nw$ an integer), and the open circles on the dashed lines indicate non-vacuum states of positive energy, constant along each dashed line.  A conventional sphaleron transition from the point A moves along such a line with equal probability in either direction. This is not the case for topological transitions along other possible paths, such as the maximal energy gradient path (on Fig.\ref{fromabove}, this path is orthogonal to the conventional sphaleron-transition path). Fig. \ref{ncsplot}, showing two cross-sections of the energy profile at $\nw =0,1$, illustrates some possible transitions which change the topological numbers and the associated changes in energy.  Fig. \ref{nwplot} shows the energy profile at constant $\ncs$, with the corrugations corresponding to tunneling barriers.

 The energy profile $E(\ncs,\;\nw =0)$ can be estimated in the case of static or slowly-changing spatially-homogeneous fields (see equation (\ref{ansatz}). For the 1+1 Abelian Higgs model \cite{grsnpb} one has:
\be\label{e11}
E(\ncsmall)/L ={1\over 2}v^2\left(gA_1\right)^2 = {3\over 2}\pi^2{\esph\over{M_H}}\ncsmall^2
\ee  
and for the present 3+1 case
\be\label{e31}
E(\ncsmall)/V = {m^4\over {2g^2}} \left((8\pi^2\ncsmall)^{4/3} + (8\pi^2\ncsmall)^{2/3}\right).
\ee
 This profile is repeated along the $\nw$
axis with a corresponding shift along $\ncs$ axis so that the minimum
$E=0$ is always on the $\ncs=\nw$ line (see Figs.~\ref{figNN},\ref{fromabove}).

As mentioned before, there is no energy barrier for movement along the
$\ncs$ axis, so this movement is completely unsuppressed and
controlled by the dynamics of the gauge field condensate $\phi$ according
to equation (\ref{e31}). This lack of suppression means that any (potentially
large) number of baryons produced by variations in $\ncs$ would
instantly disappear immediately after $\ncs$ returned to the
energetically favorable value $\ncs=\nw$, provided there were no
transitions in Higgs sector ($\nw$ remained unchanged). In other
words, even though the fermions are coupled through the anomaly to
Chern-Simons number, long-living fermionic states  appear only
due to transitions in Higgs sector which modify the energy profile
along $\ncs$ axis by moving the center (minimal energy) value of
$\ncs$ around which it oscillates.

As opposed to the absence of barriers for variations in $\ncs$ only, transitions along the $\nw$ axis
require passing over sphaleron-like barriers (for simplicity, the barriers are shown at half-integer values in Fig.~\ref{figNN}). Precisely how high this barrier is depends on several circumstances, to be discussed below, notably the Higgs VEV and the spatial scales of the gauge and Higgs fields.  In 3+1 dimensions   spatially-homogeneous fields have a high barrier (provided the Higgs VEV is not zero), by an argument similar to that given earlier:   it
would be energetically favorable to separate regions with nontrivial
gauge and Higgs fields in sphaleron configuration). In other words, the sphaleron, of size $\sim m^{-1}$, corresponds to the $minimum$ barrier height and the barrier height is greater for all other spatial scales.  However, 
in the 1+1 Abelian Higgs model  the barrier at any $\ncs,\;\nw$ is the same
as the sphaleron barrier, as can be shown   by making a large gauge
transformation that eliminates the 
background gauge field and taking into account that the gauge field
component of the 1+1-dimensional sphaleron \cite{sph11} is inversely
proportional to the spatial volume. So the contribution of the gauge field
to the (static) energy vanishes at infinite volume, while the Higgs field
contribution is identical to that of the sphaleron.  The 3+1 dimensional
case will be discussed below.   

In the homogeneous case discussed in previous sections, when the
dimensionless gauge potential $\phi$ is of order one there is both a large
Chern-Simons number and a large energy, but zero Higgs winding number
$\nw$.  It is possible that the energy in a volume containing one unit of
Chern-Simons number can exceed the usual sphaleron energy, and (as we show
in the next section) this does mean that transitions which change the Higgs
winding number become unsuppressed.  Note also that literal sphaleron
transitions, 
which change $\ncs$ and $\nw$ equally, on average do not change the Higgs
winding number, which is what is required to produce baryons. 
These do not change $E(\ncs-\nw)$ and so
their rate is not affected by the presence of a nonzero $\ncs$ density. At
any  $\ncs$, the rates of transitions with $\Delta\ncs=+1$ and 
$\Delta\ncs=-1$ equal each other and the
contribution of normal topological transitions to total
baryoproduction is zero.

\subsection{\label{efft}Effective-temperature estimates of transition rates}

Here we give some simple estimates of transition rates based on a commonly-used approximation, in which the net rate of baryon production is evaluated from a first-order expansion of the rate in powers of an energy difference, leading to a form in which the rate is evaluated as a zeroth-order sphaleron-like part times a certain energy difference.  In order that this be useful, one requires explicit knowledge of this energy dependence.  A further simplification, used here, is to assume a Boltzmann dependence with an effective temperature.  In Section \ref{rates}
we give an approximate study of the sphaleron-like part of the rate, looking for conditions under which the tunneling barrier to baryoproduction can vanish.

To evaluate the transition rate in the Higgs sector,  assume that the
transition from zero winding to $\nw=1$ proceeds along $\nw$ axis
on the path $AB$  (see Fig.~\ref{fromabove}). Then the barrier height $E_+$ along this path
can be estimated as 
\be\label{eminus}
E_+ = \esph - \left(E(\ncs) - E(\ncs-1/2)\right)
\ee
Similarly, the height of the barrier in the opposite
direction is $E_- = \esph + (E(\ncs-1/2) -
E(\ncs-1))$. If one could further assume a Boltzmann distribution
with an effective temperature $\teff$, one would obtain different rates $\Gamma_{\pm}$ for transitions in the two directions, with a net rate: 
\be\label{myrate}
\dot{B}/V=\ndw /V = \Gamma_+ - \Gamma_- \simeq {\Gamma_{\rm
sph}\over{2}}{{1}\over{\teff}}\,{{\partial E}\over{\partial\ncs}}
\ee
where $\Gamma_{\rm sph}$ is the sphaleron rate at zero energy change.
Then with the help of
equations~(\ref{e11},\ref{e31}), the winding number rate of change in 1+1 dimensions is:
\be\label{rate11}
\ndw /L = 3\pi^2 \Gamma_{\rm sph}{\esph\over\teff}{\ncsmall\over{M_H}}
\ee
and in (3+1) dimensions: 
\be\label{rate31}
\ndw /V =
{\Gamma_{\rm sph}\over\teff} {\ncsmall\over{|\ncsmall|}}
{{2\pi}\over{3\alpha_W}} 
\left(2(8\pi^2\ncsmall)^{1/3}+M_W^2(8\pi^2\ncsmall)^{-1/3}\right)
\ee
(here we assume the barrier height at $\nw=0$  equals $\esph$, and
note that the total rate is $\Gamma_{\rm sph}=\Gamma_+ + \Gamma_-$).

It is worth noting that the expression \eq{rate31} becomes infinitely large at
small $\ncsmall$. This divergence appears because in 3+1 dimensions the
homogeneous $ansatz$ \eq{smaction}--\eq{higgs} used throughout the present paper is inadequate
for field configurations with small topological charge densities,
simply because gauge field configurations with unit Chern-Simons
number are localized in space (this is not the case in 1+1 dimensions).

Strictly speaking, it is unclear whether the topological transitions in
Higgs sector proceed along the $\nw$ axis or along the maximal gradient
direction orthogonal to the $\nw=\ncs$ line. However, the energy gains
$E_{\pm}-\esph$ in the latter case are larger only by factor of 2, which
makes no qualitative difference in our analysis.

With the increase in $\ncs-\nw$ the energy gain obtained by transitions
with $\nw=\pm 1$ also increases, so at very large values of $\ncs-\nw$
transitions along the $\nw$ axis also become unsuppressed because the
energy gain \eq{eminus} becomes equal to or greater than the barrier
height, {\it i.e.}, the energy of a sphaleron-like Higgs configuration. The
critical values can be estimated by putting $E_-=0$ in \eq{eminus} and
using equations~(\ref{e11},\ref{e31}):
\be\label{crit11}
\ncsmall^{crit} \sim {2\over{3\pi^2}} M_H = {2\over{3\pi^2}} {L_{\rm
sph}}^{-1} 
\ee
In proper dimensionless units
\cite{grsnpb} equation \eq{crit11} becomes $\ncsmall^{crit} \sim 1/10$ or
$A_1\sim 1/2$) in 1+1 dimensions and 
\be\label{crit31}
\ncsmall^{crit} \sim {1\over{2\pi^2}}m^3  
\ee
in 3+1 dimensions, which is about $1/(2\pi^2)$ per sphaleron volume.
Equation~\eq{crit31} is equivalent to $\phi\gsim 1$, comparable to the
maximum amplitudes of $\phi$ seen in the numerical simulations.

\subsection{\label{baryoproduction} Baryoproduction}

Let us give some simple estimates of baryon production in the 1+1 Abelian
Higgs model.\footnote{ The estimates in sections
\ref{baryoproduction} and \ref{washout} can be straightforwardly
generalized to the 3+1 dimensional case using equation (\ref{rate31}),
which should be applicable at least for large $\ncsmall$.}  Taking into
account that the number of long-living fermionic states is equal to the net
shift of Higgs winding number, and using equation \eq{rate11}, one obtains
\be\label{bdot}
\dot{B} = 3\pi^2 {\esph\over{M_H}}{\Gamma_{\rm sph}\over\teff}\ncsmall L
\ee
and
\be\label{bint}
\Delta B = 3\pi^2{\esph\over{M_H}} \int dt\, {\Gamma_{\rm
sph}\over\teff}\ncsmall L 
\ee
where $\Delta B$ is the total number of generated baryons. 

Although the detailed discussion of $\Gamma_{\rm sph}$ and $\teff$
time dependence goes beyond the scope of this paper, there are no
reasons to expect any qualitative difference from previous studies
\cite{ggks,gg,ew2000}. There it was argued that $\Gamma_{\rm sph}/\teff$ is a
 smooth function of time with a sharp increase at the beginning of
preheating and a slow decrease in the course of thermalization to the final
reheating temperature. The exact timing depends on initial
conditions. However, equation \eq{bint} provides a simple way to check how
efficient the baryoproduction is for {\em any} reasonable time
evolution of $\Gamma_{\rm sph}$ and $\teff$. As long as the frequency
of $\ncs$ oscillations is much larger the inverse thermalization time,
we may substitute $\Gamma_{\rm sph}/\teff$ by 1 in the right-hand side of
equation \eq{bint} which turns into the integrated average of Chern-Simons
number density
\be\label{ncsavg}
\ncsmt = \int_0^t dt\, \ncsmall(t)
\ee
Time intervals when $\ncsmt$ remains constant give zero
baryoproduction; an increase or decrease in $\ncsmt$ corresponds to a
nonzero secular average of $\ncs$ and thus to
positive or negative baryoproduction (provided the transition rate
$\Gamma_{\rm sph}(t)$ is nonzero). The simplest way to get a nonzero
$\ncs$ secular average is to introduce a bias or tilt in $E(\ncs)$ 
through CP-violating terms either directly coupled to $\ncs$ in the
bosonic Lagrangian, as in equation \eq{cpop1}, or coupled to the fermion
density 
operator of equation \eq{missing} which modifies $E(\ncs)$ through the anomaly.
 It is also possible to shift the time average of $\ncs$ in other ways
through dynamical effects already described in section \ref{cpeqns}. 

\subsection{\label{washout} Washout from fermion backreaction}

Washout is one of the most prominent forms of fermionic
backreaction. As long as the topological transitions keep going, 
newly-created baryons tend to disappear through the same anomaly mechanism
that created them, to the extent that the decrease of fermionic density
reduces the system energy. Washout (in 1+1 dimensions) can 
be accounted  for  by adding a dissipative term to equation \eq{bdot}:
\be\label{bwash}
\dot{B} = 
3\pi^2 {\esph\over{M_H}}{\Gamma_{\rm sph}\over\teff}\ncsmall L - \Gamma_B B
\ee
where the dissipation rate $\Gamma_B$ is generally proportional to the
sphaleron rate $\gsph$.  In 3+1 dimensions \cite{ggks}, 
\be
\label{gammab}
 \Gamma_B = {39\over 2}{\Gamma_{\rm sph}\over{\teff^3}}
\ee
The solution of equation \eq{bwash} can be found in the form (compare to
equation (\ref{bint})): 
\be\label{fullwash}
\Delta B =  3\pi^2{\esph\over{M_H}} e^{-\gamma(t)} \int_0^t d\tau\,
e^{\gamma(\tau)} 
{{\Gamma_{\rm sph}(\tau)}\over\teff(\tau)}\ncsmall(\tau)L  
\ee
(here $\gamma(t) = \int_0^t d\tau\, \Gamma_B(\tau)$).
Direct use of equations~(\ref{bwash},\ref{fullwash}) appears to be problematic
in our simulations for two reasons. First, explicit time dependences
of the rate $\Gamma_{\rm sph}$ and the effective temperature $\teff$ are
controlled by dynamics in the inflaton and Higgs sector which are beyond
the scope of the present paper. Second, equation \eq{bwash} assumes the
fermions to be 
thermalized, which obviously is not the case if Chern-Simons number
oscillates with frequency of order of $m$.

However, it is easy to see that solution \eq{fullwash} decays
exponentially with time if $\ncsmall=0$ and $\Gamma_B \neq
0$. Therefore, it is desirable to keep baryoproduction going until the
topological transitions end (e.g. because of freeze-out during
thermalization to a low reheating temperature). Otherwise, if
baryoproduction ends before the sphaleron transition rate approaches
zero, wash-out could crucially affect the final density of baryons. 

Again, the time dependence of the $\ncsmall$ running average \eq{ncsavg}
provides  important information about the baryoproduction period and
allows one to estimate acceptable thermalization times.  For example,
short-term baryoproduction followed by stabilization of $\ncsmt$ at a
certain value (see Fig.~\ref{cp1fig2}) would seldom survive the
wash-out, while steady baryoproduction (such as is shown in
Fig.~\ref{cp1fig3}) leaves ample time for the sphaleron transitions
to freeze.

\subsection{\label{rates} Dependence of tunneling barriers on $\phi$ and
scale sizes}

Our purpose here is to find approximately the conditions under which the
tunneling barrier in the sphaleron rate $\Gamma_{\rm
sph}=\Gamma_++\Gamma_-$ (see equation 
(\ref{myrate})) can vanish.  It will not be necessary to use an
effective-temperature approximation.  The techniques used here, based on
work of a decade or more ago \cite{bc,co89,cr,mamo} on 
topological charge-changing transitions, are qualitative but useful.  They
go  beyond 't Hooft's original work, 
whose famous tunneling factor of $\exp [-8\pi^2 /g^2]$ holds only at zero 
energy and, at that energy, is (classically) independent of scale size.  We
give an approximate barrier-factor formula with explicit dependences on
energy, scale size, and Higgs VEV, and outline the regions of this
parameter space where the tunneling exponent can vanish.  We do not discuss
the origin or distribution of spatial scales, but simply assumes that these
arise through some mechanism such as the unstable growth of spatial ripples
\cite{alexmike}.   

There are other interesting approaches to this kind of problem, which we
intend to investigate in the future.  These include studies of non-Abelian
gauge dynamics in the presence of a non-vanishing topological charge
density \cite{niki} or in the presence of a time-varying electric
background potential \cite{ndg}, and a study of the conversion of a
time-varying background topological charge to fermions through the anomaly
\cite{kl}.  These works use various specialized backgrounds not exactly
comparable to the Chern-Simons condensate used in the present paper, but
they should still have qualitative applicability. 

There are at least  two mechanisms which can remove tunneling barriers, and
these may operate at the same time. The first mechanism is generally
important for large scale sizes (large means compared to the vacuum W-boson
mass); it involves swinging of the Higgs VEV through zero.  The second
mechanism involves selection of a spatial scale such that the original
Chern-Simons condensate is well-matched to baryon production through the
anomaly.  As one might expect, the best overlap occurs when the size scale
is about $m^{-1}$, and the corresponding configurations are somewhat like
sphalerons in size, but it is apparently possible for tunneling barriers to
be overcome at considerably larger spatial scales.  However, baryon
production is inefficient at these large scales.

First we discuss the conditions for no tunneling barriers with a fixed and
finite W-boson mass $m$, and then remark briefly on what  happens when
there is no mass because the Higgs VEV vanishes. 
 
\subsubsection{The no-barrier condition for finite W-boson mass}

There must be a transformation of Chern-Simons number from the
spatially-homogeneous condensate to one characterized by spatial
inhomogeneities such as sphalerons.  That is, in the expression for
$\ncs$
\begin{equation}
\label{ncs}
\ncs=\frac{-1}{8\pi^2}\int
d^3x\epsilon_{ijk}{\rm\,Tr\,}[A_i\partial_jA_k+\frac{2}{3}A_iA_jA_k] 
\end{equation}
the relevant sphaleron-like configurations of the $A_i$ are typified by
gauge potentials which approach at infinity pure-gauge terms carrying
winding number; the Higgs field carries the same winding number for a
minimum-energy configuration.  The gauge in question was termed $U$ in
equations (\ref{gauge}, \ref{higgs2}).

A typical form for a gauge matrix $U$ carrying unit winding number is
\cite{bc,co89}: 
\begin{equation}
\label{bc}
U=\exp \left[\frac{i}{2}\vec{\tau}\cdot \hat{r}\beta (r,t)];\;\;\beta
(r,t)=2\arctan [r/\lambda (t)\right] 
\end{equation}
where $\lambda (t)$ is a monotonic function of $t$ going to $-\infty$ at
$t=-\infty$ and to $\infty$ at $t=\infty$.  A gauge matrix for Chern-Simons
number $\ncs$ is simply a product of $\ncs$ terms of the form (\ref{bc})
translated to various spatial and temporal centers.  
We will refer to the spatially-homogeneous gauge potentials dealt with in
earlier sections as $\phi$ form potentials, and the spatially-dependent
sphaleron-like configurations as $U$ form potentials.  Of course, either
form is at best an approximation; the $\phi$ form potentials will develop
spatial gradients by several mechanisms,  and the $U$ form potentials will
not literally be of the approximate form we use below.  

Long ago, Bitar and Chang \cite{bc} constructed Minkowski-space gauge
potentials of unbroken $SU(2)$ gauge theory  whose asymptotic behavior was
precisely that of equation (\ref{bc}).  They chose the single dynamical
degree of freedom to be the function $\lambda (t)$ of this equation; the
non-asymptotic gauge potential is further parameterized by  a non-dynamical
scale coordinate and translation coordinates.  The parameterization is
conveniently written as: 
\begin{equation}
\label{bcpot}
A_{\mu}= \left(\frac{r^2+\lambda^2}{r^2+
\lambda^2+\rho^2}\right)U\partial_{\mu}U^{-1}    
\end{equation}
where $\rho$ is the scale coordinate (we will not need to display the
translation coordinates, which we set to zero).  Bitar and Chang \cite{bc}
show that the topological charge of this potential is unity for any scale
coordinate value.  They also show that the dynamical degree of freedom
$\lambda$ has a Hamiltonian which is quadratic in $\dot{\lambda}$ but a
complicated function of $\lambda$, and computed the barrier exponent $\int
pdq$ at zero energy from this Hamiltonian, recovering the 't Hooft result.
Later this work was extended \cite{co89} to EW theory with a Higgs field
(as well as a chemical potential for Chern-Simons number); this work then
formed the basis \cite{co90} for an investigation of scattering processes
involving topological charge change at very high energies.  In this paper,
we extend these earlier works to cover the entire range of scale
coordinates (Ref. \cite{co90} only covered the regime of small scale
coordinates).

We wish to find the conditions under which the corrugations or barrier
factors described in connection with Fig. \ref{figNN} vanish, within a
framework general enough to go beyond thermal quasi-equilibrium.  
Begin with a general formula of Cline and Raby \cite{cr} for the diffusive
rate.  Originally the Cline-Raby formula was given for thermal equilibrium
conditions, but it is easily modified for non-equilibrium conditions.  The
derivation is slightly different from Cline and Raby's because of the
non-equilibrium nature of the process.  Consider diffusive dynamics of the
Chern-Simons number $\ncs(t)$ in which the diffusion
constant\footnote{Roughly speaking the diffusion constant $\Gamma$ used
here is equivalent to $V\gsph$ of section \ref{efft}.} $\Gamma$ 
can be written in the usual form:
\begin{equation} 
\label{diff} 
\frac{\Gamma}{V}=\frac{1}{2V} \int^{\infty}_{-\infty}\langle
\ndcs(t)\ndcs(0)\rangle=\frac{\pi}{V} \sum_{if}\rho(i)\delta
(E_f-E_i)|\langle i,E_i|\ndcs(0)|f,E_f\rangle |^2  
\end{equation} 
where $V$ is the volume of space and the brackets refer to a trace over the
density matrix; this density matrix is, for the present purposes, taken to
be diagonal in the energies $E_i,E_f$ of the states summed over, with
entries $\rho(i)$. In fact, the density matrix is changing in time and so
does not commute with the Hamiltonian, but at the present level of
(in)accuracy this is an inessential complication.  Cline and Raby  take the
$i$ states to be in states at $t=-\infty$ and the $f$ states to be out
states at $t=+\infty$, and then use the formula:  
\begin{eqnarray} 
\label{creq} 
\int^{\infty}_{-\infty}dt\langle i,E_i|\ndcs(t)|f,E_f\rangle  =  2\pi\delta
(E_i-E_f)\langle i,E_i|\ndcs(0)|f,E_f\rangle \\ \nonumber  
=\langle i,E_i|\ncs(\infty )-\ncs(-\infty )|f,E_f\rangle  =  i\Delta
N_{fi}(2\pi)^4\delta_4(p_f-p_i)T_{fi}  
\end{eqnarray} 
where $T_{fi}$ is the zero-temperature S-matrix element from the initial to
the final state, and $\Delta N_{fi}$ is the change in Chern-Simons number
(or topological charge) from the initial to the final state.  Substitution
in (\ref{creq}) yields:  
\begin{equation} 
\label{crsecond} 
\frac{\Gamma}{V}=\frac{1}{2}\sum_{if}\rho(i)(2\pi)^4\delta_4(p_f-p_i)( 
\Delta N_{fi})^2|T_{if}|^2. 
\end{equation}

Further progress depends on analysis of B+L-violating S-matrix elements at
zero temperature, a subject of some considerable interest a decade ago
(see, {\it e.g.}, \cite{mamo}).      Ref. \cite{co90} gives the following
very crude approximation to the  $|\Delta N_{fi}|=1$ S-matrix elements at
fixed energy $E$:  
\begin{equation} 
\label{c90} 
T_{if}\sim (\frac{4\pi^2}{g})^N\int_0^{\infty}d\rho \rho^{N-5}\prod_j
e^{-\rho k_j}e^{-Q(\rho ,E)}  
\end{equation} 
where $Q$ is a barrier exponent, $N$ is the total number of particles
involved in the scattering process,  $\rho$ is a scale collective
coordinate, and $k_j$ is the magnitude of the three-momentum of particle
$j$.   This formula is based on a transcription of familiar Euclidean
formulas for scattering in the presence of instantons to Minkowski space.
Unlike naive instanton-based amplitudes, the above amplitudes $T_{if}$
behave properly at high energy (where one expects $\rho \simeq N/E$) but
are not correctly  unitarized; this will not be an important issue
here.\footnote{See \cite{hc} for a multi-channel study of unitarization
effects.}

The barrier exponent $Q$ was originally \cite{co90} given for small $\rho$.
The appropriate expression for all $\rho$ can straightforwardly found using
the techniques of \cite{bc,co89}:  
\begin{equation} 
\label{barrier1} 
Q(\rho , E)=\frac{6\pi^2}{g^2} \int_{-\infty}^{\infty}d\xi
F(\xi)^{1/2}\Theta [F(\xi)], 
\end{equation} where the function $F$ is:
\begin{equation}
\label{barrier2}
F(\xi)=(\xi^2+1)^{-5/2}[(\xi^2+1)^{-5/2} +m(t)^2\rho^2f(\xi
)-\frac{1}{3\pi^2}E\rho g^2]. 
\end{equation}
Here we make it explicit that the W-boson mass $m(t)$ depends on time,
because the Higgs field depends on time. 
The function $f(\xi )$ is not expressible analytically, but one can show
that $f$ is positive and obeys $f(\xi)\leq f(0)=4/3$.  

The barrier factor $Q$ vanishes if the function $F$ is always negative,
which happens for certain regimes of energy $E$, mass $m(t)$,  and scale
coordinate $\rho$.  Clearly, there is always a finite barrier at $\rho=0$
(just the 't Hooft barrier if $m$=0).  For non-zero $\rho$  it is generally
true that if $F(\xi =0;\rho, m, E)$ vanishes then $F\leq 0$ for all $\xi$.
The no-barrier condition then is:  
\begin{equation}
\label{nobarrier}
1+\frac{4m^2\rho^2}{3}-\frac{E\rho g^2}{3\pi^2}\leq 0.
\end{equation}
Consider first the usual case where $m$ is the standard W mass.
The minimum energy $E$ yielding equality in equation (\ref{nobarrier})
occurs at $\rho =\sqrt 3/2m,\;E_{min}=4\pi^2 \sqrt 3m/g^2$.  This minimum
should be the sphaleron energy, and it is indeed a very good numerical
approximation \cite{co89} in the limit of large Higgs mass. 

Next we take up the case where the W mass, or Higgs VEV, is near zero.

\subsubsection{Higgs VEV near zero}

It appears that large scales $\rho$, which would be expected in the first
stages of transition of Chern-Simons number from $\phi$ form  to $U$ form,
will have a disastrously large barrier factor going like $\exp (-{\rm
const}\;m\rho /g^2)$, if the mass $m$ is anywhere near its standard value.
However, if---as discussed in Sec. 5A---the Higgs field $v$ oscillates
through zero because of gauge back reaction, the gauge mass $m=gv/2$
vanishes and the barrier $Q$ will vanish periodically (see (\ref{barrier2})
at energies $E\sim 1/(\rho g^2)$; that is, more easily at large scales than
at small.\footnote{The idea that preheating causes oscillations or
vanishing of $v$ and therefore reduction or elimination of the barrier
factor is given in Refs. \cite{gg,kof}.}  So Higgs oscillations are one
potentially-vital means of generating baryons from Chern-Simons number.  It
is easy to check that if the barrier factor is unity for a time $\tau$
during a Higgs oscillation period $\tau_{\rm Higgs}$, the averaged barrier
factor over a Higgs period is not exponentially small, but is of order
$\tau /\tau_{\rm Higgs}$.  
  
Unfortunately, this is not the end of the story, since there are other
possible suppression effects to deal with; a poor overlap of
anomaly-produced baryons with the Chern-Simons condensate can be
disastrous.   

\subsection{Good overlap condition}

Here we explore how the conditions of both no barrier as well as a good
overlap between a $U$ form potential and baryons produced through the
anomaly can be satisfied.

Even if $v$ does not vanish during preheating it is possible in principle
to have a zero barrier factor, depending on what values of the
dimensionless gauge potential $\phi$ are reached. Values of $|\phi |$ near
or slightly greater than one are fairly readily gotten, as is evident from
the structure of the equations of motion and from the numerical studies
reported above, and for such values of $\phi$ there need be no barrier.
However, it is possible for $|\phi |$ to be small compared to unity; we
explore that possibility here.  As one might expect, for values of $\phi$
rather less than one, it requires a rather large region to gather together
enough energy to overcome the barrier. It will turn out that spatial scales
much larger than the inverse of the vacuum W-boson mass are self-consistent
only if the Higgs VEV does go near zero. 

Consider a region of space of size $\rho$ as defined by gradients appearing
during preheating and further unstable amplification.  In this region the
total energy and Chern-Simons number are approximately: 
\begin{equation}
\label{totalen}
E(\rho)\simeq \frac{4\pi m}{g^2}(m\rho )^3(\phi^2+\phi^4);\;\;N_{tot}(\rho
)\simeq \frac{(m\rho \phi )^3}{6\pi}. 
\end{equation}
Inserting these estimates in the no-barrier condition (\ref{nobarrier}) yields:
\begin{equation}
\label{nobarrier2}
(m\rho
)^2=
\frac{\pi}{2(\phi^2+\phi^4)}\left\{1+[1+\frac{3}{\pi}(\phi^2+
\phi^4)]^{1/2}\right\}\simeq  
\frac{\pi}{\phi^2}, 
\end{equation}
with the last form holding for small $|\phi |$.  In this case the total
energy and Chern-Simons number in terms of $\phi$ are: 
\begin{equation}
\label{encs}
E=\frac{4\pi^{5/2}m}{g^2|\phi |};\;N_{tot}=\frac{\pi^{1/2}}{6}=O(1).
\end{equation}

The appearance of inverse powers of $\phi $ suggests the inefficiencies of
avoiding a barrier when parametric resonance amplification of the EW gauge
potential is small:  The energy is large compared to the sphaleron mass,
but the change in B+L is only of order one (per unit flavor).  Still, these
inefficiencies could be tolerable in view of the exponentially-small
efficiency of actual barrier penetration.  When amplification leads to
$|\phi |\sim 1$, avoiding a barrier is rather like having energy at or
above the sphaleron mass. 

Next turn to the conditions specifying a good overlap between the $\phi$
condensate and baryogenesis.  We will find that a good overlap means $|\phi
|$ cannot be too far below unity, but we cannot quantify this statement
with the present crude approximations.  The point of a good overlap is that 
having an energy larger than the barrier energy is by no means sufficient
in many cases to lead to unsuppressed B+L violation, as many authors have
discussed \cite{mamo,co90}.  For example, in the formula (\ref{c90}) for
S-matrix elements, the other factors integrated over $\rho$ in that
equation can lead to suppression of the S-matrix element by a factor of
order $\exp [-\zeta 8\pi^2/g^2]$ with $\zeta\sim 1/2$  or so \cite{co90}.
This sort of suppression even in the absence of a tunneling barrier comes
about because of a very poor overlap between multi-particle initial and
final states when these have very different numbers of particles and at
least one particle number is very large.  When the states are sufficiently
similar there is no such extra suppression, which is what happens in
thermal equilibrium at large enough temperature.  In the present case
something similar happens.  The initial state in the Cline-Raby formula
(\ref{creq},\ref{crsecond}) is not a conventional particle state; it is a
coherent state somewhat similar to the $\phi$ form potential, but with a
spatial size $\rho$ coming from various effects, such as growth of unstable
momentum modes \cite{alexmike}.  If this state has a large overlap with the
$U$ form potentials such as in (\ref{bcpot}), {\em and} if there is no
barrier for tunneling, then the amplitude for baryon creation will be
unsuppressed.  One can say that the exponentially-small rate of B+L
violation in $2\rightarrow N$ collisions stems from the Drukier-Nussinov
effect \cite{dn} that it is extremely unlikely for a two-particle collision
state to couple well to a soliton like the sphaleron, but in our case the
Chern-Simons condensate may, for certain values of $\phi$ and $\rho$,look
enough like a ``soliton" for there to be good overlap.  

We seek this substantial overlap between a generic $\phi$ form potential,
somehow modified to have an overall spatial scale $\rho$, and a $U$ form
(Bitar-Chang) potential, when $\rho$ is a large scale compared to other
spatial scales.  At large $\rho$ the Bitar-Chang fields scale as: 
\begin{equation}
\label{bcfields}
gE\sim \frac{\dot{\lambda}}{\rho^2};\;gB\sim \frac{1}{\rho^2}.
\end{equation}
By comparison\footnote{The fields $\phi ,\;\dot{\phi}$ in equation
(\ref{EB}) are dimensionful; the ones we use now are scaled by appropriate
powers of $m$ to be dimensionless.}to the $\phi$ form potentials of
equation (\ref{EB}) one sees from the B field consistency with the relation
$m\rho \sim 1/|\phi |$, and that consistency with the E field can be
achieved if $\lambda \sim \ln \phi$.   

Now return to the approximate form of the S-matrix elements (\ref{c90}).
The process to be described is the transformation of a $\phi$ form field
with certain spatial scales to a final state of $N$ particles, including a
set of particles which violates B+L.  The process is only interesting if
there is no tunneling barrier, so we assume the validity of equation
(\ref{nobarrier}).  The next question to ask is whether one can, consistent
with  (\ref{nobarrier}), argue for a non-suppressed overlap between the
modified $\phi$ form and the final state.  In the crude approximation of
equation (\ref{c90}) this simply amounts to asking whether the constraint
on scale coming from the final-state wave function is consistent with other
information on that scale, such as the condition (\ref{nobarrier}) for no
barrier.  In (\ref{c90}) there occurs a product of exponentials of the type 
$\exp (-k_i\rho )$ where $k_i$ is the spatial momentum of the $i^{th}$
particle.  It is reasonable to assume that all the particles are
effectively massless, and then the product of wave functions reduces to
$\exp (-E\rho )$.  One easily sees that, if the barrier exponential factor
$\exp -Q$ is unity, the $\rho$ integrand maximizes at $\rho=N/E$, just as
estimated \cite{co90} for more conventional S-matrix elements.  So a good
overlap simply means that the number of produced particles is (more or
less) determined.  This turns out, taking into consideration the no-barrier
condition (\ref{encs}), to yield $N\sim 1/g^2\phi^2$.  (One might argue
that in fact particles do have mass, and so one should have $N\leq E/M\sim
1/g^2|\phi|$ which would require $|\phi|$ at least of order unity.
However, $m$ is the W-boson mass, and all baryons and leptons except for
baryons with top quarks are not nearly that heavy.)  

As in \cite{co90} we can form a rate from the Cline-Raby formula
\ref{crsecond} by multiplying the squared S-matrix elements from
(\ref{c90}) by massless $N$-particle phase space \cite{co90}.  Nothing
quantitative should be trusted about the resulting equation
(\ref{gammarate}) below, except for its dependence on $\phi$: 
\begin{equation}
\label{gammarate}
\frac{\Gamma }{V}\sim \frac{E^4}{N!g^{2N}}\sim
(\frac{m}{g^2\phi})^4(\frac{\phi^2}{\phi_c^2})^N. 
\end{equation}
Here $\phi_c$ is a critical value of $\phi$ separating small from large
rates (in the present highly inaccurate approximation, $\phi_c^2=e^{-1}$).
Whatever $\phi_c$ is, and it is evidently of order unity, it is clear that
for $N\geq 2$ and for $\phi^2\geq \phi_c^2$ the B+L process is
unsuppressed, while in the opposite limit it is strongly suppressed.  Note
that if the dimensionless potential $\phi$ is of order unity the energy
scale is the sphaleron mass and the spatial scale is $m^{-1}$.  So we
conclude, not unexpectedly, that if the Higgs VEV is not oscillating near
zero spatial scales near the sphaleron mass must be formed; if the Higgs
field is oscillating near zero, much larger spatial scales will serve for
unsuppressed B+L production.    

\section{Conclusions} 
 
In this work we studied a scenario of baryogenesis in the Standard
Model, based on inflation on EW scales with the Higgs coupled to the
oscillating inflaton in a preheating phase.  If baryons could be
generated, they could be saved from washout at reheating because the
reheat temperature is less than the EW cross-over temperature where
sphaleron washout can be large,  so the topological transitions can
completely freeze out before the preheating ends. However, the
dynamics of conversion of Chern-Simons condensate to B+L and the
dynamics of washout of this B+L are very complex and not yet
well-understood.

Although we have no rigorous proof that  B+L production involving only
standard-model fields (coupled to the inflaton) is truly viable,  we can at
least say that so far that we have not identified any mechanisms which rule
out a pure EW/inflaton scenario.

Extending the earlier work of Ref. \cite{alexmike}, we added explicit
CP violation to the spatially-homogeneous classical gauge equations of
motion, rather than (as in \cite{alexmike}) depending only on initial
value of the functions 
$\phi ,\dot{\phi}$ for CP violation.  The effects of explicit
CP-violating terms was studied analytically in some simple models, as
well as numerically.  We studied numerically the classical dynamics of
the Higgs field, including classical back reaction of the gauge field
on the Higgs field.

The gauge reaction on the Higgs equation of motion is important in two
respects: It can broaden parametric resonances by getting away from pure
sinusoidal variation of the Higgs field, which was the only case considered
in \cite{alexmike}.  And it can, as discussed above, lead to conversion of
Chern-Simons number to baryons which for some fraction of the time is
unsuppressed by tunneling barriers, since gauge backreaction often modifies
the Higgs potential and causes the Higgs field to oscillate through zero,
rather than staying at the bottom of the potential well.
      
There are various forms of CP violation which could be added to the
equations of motion; we explored three in the gauge equations and one in
the Higgs equation.   In the gauge sector one  came from strong
out-of-equilibrium CP violation; another came from a multi-Higgs sector
with spontaneous CP violation; and the third was a higher-derivative
CP-violating term, which could be associated with other lower-dimension
operators.  In the Higgs sector we explored spontaneous baryogenesis in a
multi-Higgs model, leading to an effective chemical potential for baryons.
Generally speaking, the resultant secular average of the Chern-Simons
number is a few orders of magnitude less than the dimensionless coefficient
multiplying the CP-violating term in the gauge equations of motion. 

Formation of a spatially-homogeneous Chern-Simons condensate is only the
beginning of the story; it is necessary to convert this condensate into
something resembling a condensate of sphalerons of the same Chern-Simons
number, in order to generate the Higgs winding number which  is converted
to baryons.   The usual approach of invoking a thermal (or effectively
thermal) regime immediately following the parametric resonance regime is
not likely to be applicable.  A quantitatively-adequate study of these
non-thermal non-equilibrium processes  remains to be done, but we have
given useful criteria for evading the two possible process which lead to
exponential suppression of baryogenesis.  The first, of course, is the
topological-charge tunneling barrier.  The second is a poor overlap between
initial and final states in a process even with no barrier.    
Based on earlier work which studied the (im)possibility of baryon
production in accelerator collisions, we gave  crude estimates of the
parameter ranges which avoid both barriers and bad overlap, and pointed out
the influence of the Higgs VEV passing near zero for these parameter
ranges.  The conclusion is that if the Higgs VEV stays near its vacuum
value, spatial scales near the vacuum inverse W-boson mass and energies
near the vacuum sphaleron mass can avoid both forms of suppression, quite
unlike the $2\rightarrow N$ collision case where poor overlap cannot be
avoided.    If this VEV goes near zero, the spatial scales can be larger
and the energy scales smaller. 

If this view proves to be correct, the rate-limiting step in baryogenesis
will be the conversion of $\phi$ form EW fields to $U$ form fields, by
growth of perturbations at various spatial scales.

The scenario of regions of almost spatially-homogeneous Chern-Simons
condensate of order of the EW Hubble size, may have consequences beyond B+L
generation.  It has been proposed \cite{co97} that any mechanism of B+L
generation involving the EW anomaly will leave its trace on the early
universe through large-scale helicity of the Maxwell magnetic fields which
descend from EW gauge fields as the universe cools.  As is well-known, if
these magnetic fields have typical EW scales of $10^{-16}$ cm, the
present-day magnetic-field scale will be far too small even taking
expansion of the universe into account.  Following earlier studies in
magnetohydrodynamics, Ref. \cite{co97} proposed that the helicity would
drive an inverse cascade to longer scales.  This idea was further pursued
by Son and by Field and Carroll \cite{son}.  If the EW fields which lead to
Maxwell magnetic fields are generated on the EW Hubble size scale, one
gains many orders of magnitude toward seeding present-day intergalactic
magnetic fields by EW processes.  Moreover, there might be much more
Maxwell helicity than the minimum required to produce the present-day
abundance of baryons, since only a fraction of the $\phi$ form Chern-Simons
condensate will be turned into baryons.  The Maxwell helicity produced
during preheating amplification of CP violation can be very much greater
than the number of baryons.  Work is underway to investigate these points. 

\newpage

\begin{center}
{\large \bf Acknowledgments}
\end{center}    
This work was supported in part by the US Department of Energy, grant
DE-FG03-91ER40662, Task C, and by the NATO Collaborative Linkage Grant
PST.CLG.976397.  D. G. thanks the Elementary Particle Theory group of UCLA,
where this work was performed, for their kind hospitality during his
visit; his work is supported in part by CRDF grant
RP1-2103.  D. G. is indebted to N. Manton, S. Nussinov and M. Shaposhnikov for
discussions and comments.

\end{document}